\documentclass{emulateapj}

\usepackage{graphicx}
\usepackage[]{natbib}
\usepackage{placeins}
\usepackage{lineno}
\usepackage{ulem}

\newcommand{\degree}{\ensuremath{^\circ}}

\shorttitle{Time-Resolved Spectroscopy of the 3 Brightest and Hardest Short Gamma-Ray Bursts}
\shortauthors{Guiriec et al.}

\begin{document}

\submitted{Accepted for publication in the Astrophysical Journal September, 23 2010 (Submitted May, 16 2010)}
\title{Time-Resolved Spectroscopy of the 3 Brightest and Hardest Short Gamma-Ray Bursts Observed with the {\it FGST} Gamma-ray Burst Monitor}


\author{Sylvain Guiriec\altaffilmark{1}, Michael S. Briggs\altaffilmark{1},  Valerie Connaugthon\altaffilmark{1}, Erin Kara\altaffilmark{2}, Fr\'ed\'eric Daigne\altaffilmark{3}, Chryssa Kouveliotou\altaffilmark{4}, Alexander J. van der Horst\altaffilmark{5}, William Paciesas\altaffilmark{1}, Charles A. Meegan\altaffilmark{6}, P.N. Bhat\altaffilmark{1}, Suzanne Foley\altaffilmark{7}, Elisabetta Bissaldi\altaffilmark{7}, Michael Burgess\altaffilmark{1}, Vandiver Chaplin\altaffilmark{1}, Roland Diehl\altaffilmark{7}, Gerald Fishman\altaffilmark{4}, Melissa Gibby\altaffilmark{9}, Misty Giles\altaffilmark{9}, Adam Goldstein\altaffilmark{1}, Jochen Greiner\altaffilmark{7}, David Gruber\altaffilmark{7}, Andreas von Kienlin\altaffilmark{7}, Marc Kippen\altaffilmark{10}, Sheila McBreen\altaffilmark{11}, Robert Preece\altaffilmark{1}, Arne Rau\altaffilmark{7}, Dave Tierney\altaffilmark{11} and Colleen Wilson-Hodge\altaffilmark{4}}

\altaffiltext{1}{University of Alabama in Huntsville, NSSTC, 320 Sparkman Drive, Huntsville, AL 35805, USA}
\altaffiltext{2}{Department of Physics \& Astronomy, Barnard College, Columbia University, New York, NY 10027}
\altaffiltext{3}{Institut d’Astrophysique de Paris – UMR 7095 Universit\'e ́Pierre et Marie Curie-Paris 06; CNRS 98 bis bd Arago, 75014 Paris, France}
\altaffiltext{4}{Space Science Office, VP62, NASA/Marshall Space Flight Center, Huntsville, AL 35812, USA}
\altaffiltext{5}{NASA Postdoctoral Program Fellow, NASA/Marshall Space Flight Center, 320 Sparkman Drive, Huntsville, AL 35805}
\altaffiltext{6}{Universities Space Research Association, NSSTC, 320 Sparkman Drive, Huntsville, AL 35805, USA}
\altaffiltext{7}{Max-Planck-Institut f\"ur extraterrestrische Physik (Giessenbachstrasse 1, 85748 Garching, Germany)}
\altaffiltext{9}{Jacobs Technology}
\altaffiltext{10}{Los Alamos National Laboratory, PO Box 1663, Los Alamos, NM 87545, USA}
\altaffiltext{11}{University College, Dublin, Belfield, Stillorgan Road, Dublin 4, Ireland}

\email{sylvain.guiriec@nasa.gov, sylvain.guiriec@lpta.in2p3.fr}





\begin{abstract}

From July 2008 to October 2009, the Gamma-ray Burst Monitor (GBM) on board the Fermi Gamma-ray Space
Telescope ({\it FGST}) has detected $~$320 Gamma-Ray Bursts (GRBs). About 20\% of these events are classified as short based on their $T_{90}$ duration below 2 s. We present here for the first time time-resolved spectroscopy at timescales as short as 2 ms for the three brightest short GRBs observed with GBM. The time-integrated spectra of the events deviate from the Band function, indicating the existence of an additional spectral component, which can be fit by a power-law with index $\sim-1.5$. The time-integrated $E_{\rm peak}$ values exceed 2 MeV for two of the bursts, and are well above the values observed in the brightest long GRBs. Their $E_{\rm peak}$ values and their low-energy power-law indices ($\alpha$) confirm that short GRBs are harder than long ones. We find that short GRBs are very similar to long ones, but with light curves contracted in time and with harder spectra stretched towards higher energies. In our time-resolved spectroscopy analysis, we find that the $E_{\rm peak}$ values range from a few tens of keV up to more than 6 MeV. In general, the hardness evolutions during the bursts follows their flux/intensity variations, similar to long bursts. However, we do not always see the $E_{\rm peak}$ leading the light-curve rises, and we confirm the zero/short average light-curve spectral lag below 1 MeV, already established for short GRBs. We also find that the time-resolved low-energy power-law indices of the Band function mostly violate the limits imposed by the synchrotron models for both slow and fast electron cooling and may require additional emission processes to explain the data.
Finally, we interpreted these observations in the context of the current existing models and emission mechanisms for the prompt emission of GRBs.

\end{abstract}






\section{Introduction}

The bimodal distribution of Gamma-Ray Burst (GRB) durations, with a minimum around 2 s, was first clearly established by~\citet{Kouveliotou:1993} using data from the Burst And Transient Source Experiment ({\it BATSE}) on-board the Compton Gamma-Ray Observatory ({\it CGRO}). 
Further evidence for two separate populations was a tendency identified in the same data for the short 
bursts to have harder spectra than the long bursts~\citep{Kouveliotou:1993,Paciesas:2003}. A natural consequence of these observations 
was the speculation that the separate populations might have distinct physical origins. Since then, detailed multi-wavelength 
observations of GRB afterglows have supported a physical picture in which long GRBs are produced by the gravitational collapse of very 
massive stars~\citep{Woosley:1993,MacFadyen:1999,Woosley:2006} and short GRBs are the result of mergers of a neutron star with another neutron star or a black hole \citet[BH~--~][]{Paczynski:1986,Rosswog:2003}; alternate scenarios include the merger of a BH with a white dwarf~\citep{Fryer:1999} or a helium star~\citep{Fryer:1998}. Finally, short GRBs could also be produced by a variant of the collapsar model~(\citet{Zhang:2003}~--~for extensive reviews on long and short GRBs see \citet{Nakar:2007} or \citet{Piran:2004}).

Additional observational evidence differentiating the burst classes came from spectral lags, estimated from cross-correlations of burst 
time-profiles between different energy bands within the $25-1000$ keV {\it BATSE}/Large Area Detectors (LADs) energy range~\citep{Norris:2000}. The distributions of spectral lags of short and long GRBs differ significantly. For long GRBs the typical lag is $\sim$100 ms~\citep{Norris:2002,Hakkila:2007} with the hard photons arriving earlier (hard-to-soft evolution), whereas short GRBs 
typically have lag values consistent with zero~\citep{Norris:2006,Yi:2006}. Interestingly, however, recent results with the Fermi Gamma Ray Space Telescope ({\it FGST}) indicate that in the GeV range, photons always arrive later than the prompt gamma-rays ($10-1000$ keV), thus reversing the trend to a soft-to-hard evolution both in long and in short GRBs~\citep{Abdo:2009:GRB080916C,Abdo:2009:GRB090902B,Ackermann:2010:GRB090510}.

Whether the nature of the central engine differs between the two classes is not clear. However, the general picture of a collimated ultra-relativistic outflow seems to apply in both cases and theories for generating the prompt and afterglow emission generally do not 
distinguish between them~\citep{Piran:2004}. Although time-resolved spectral analysis may provide significant theoretical constraints, most short GRBs are 
too weak for detailed spectral analysis and one is usually restricted to simplified parametrizations such as hardness ratios and 
spectral lags, which provide rather crude constraints for theoretical models. Prompt GRB spectra can generally be fit adequately with 
the Band GRB function~\citep{Band:1993}, which is basically two power laws joined smoothly at a break energy,  $E_{\rm peak}$,  indicating the maximum of the ${\nu}F_{\nu}$ spectrum. However, the high-energy power-law is often not well measured or constrained, especially for weak and/or short bursts, and a single power-law with a high-energy cutoff can be used instead.~\citet{Paciesas:2003} investigated the distribution of the spectral 
parameters of {\it BATSE} GRBs versus duration and found that short GRBs differ from long GRBs by having both higher $E_{\rm peak}$ values and 
harder low-energy power-law indices.~\citet{Ghirlanda:2004,Ghirlanda:2009b} studied a sample of relatively bright {\it BATSE} GRBs and found 
that the hardness difference was attributable mainly to the low-energy power-law index with no difference in~$E_{\rm peak}$~values. They also compared the short-GRB spectral-parameters with the first few seconds of long GRBs and found no significant difference.

The Gamma-ray Burst Monitor (GBM) on board {\it FGST} has been detecting GRBs since July 2008. From July 2008 to October 2009, GBM triggered on 
about 320 GRBs, of which approximately 20\% have durations shorter than 2~s (Paciesas et al., in 
preparation). With the unprecedented time resolution of GBM, down to about 2 $\mu$s, we have the ability for the first time to perform fine-time resolved spectroscopy of the brightest short GRBs at a millisecond timescale. Such detailed analysis is a key point in understanding 
the differences and similarities in the behavior of the two classes of GRBs and their emission 
mechanisms.

In this paper we have selected a subset of the three brightest short events detected with GBM and performed detailed 
temporal and spectral analyses at millisecond timescales. In section~\ref{section:GBM}, we describe the instrument, pointing out its assets in the context of the following analysis, as well as the data set we used. Section~\ref{section:observation} is devoted to the timing analysis with time-lag and duration studies. The time-integrated and time-resolved spectral analysis of the three selected events are presented in section~\ref{section:spectral-analysis}. Section~\ref{section:interpretation} is dedicated to the theoretical interpretation of our observations. 


\section{Instrumentation and Data}
\label{section:GBM}

{\it FGST} was launched on 2008 June 11 with two instruments on board: the Large Area Telescope (LAT) and GBM. The latter instrument is an array of scintillation detectors sensitive to gamma-rays in the energy range 8 keV to 40 MeV. Twelve sodium iodide (NaI) detectors, 12.7 cm in diameter by 1.27 cm thick, cover energies up to 1 MeV and are used to determine 
burst locations. Two bismuth germanate (BGO) detectors, 12.7 cm diameter by 12.7 cm thick, cover energies above 200 keV. The high 
sensitivity of GBM above 1 MeV is ideal for the study of hard GRBs. Moreover, the broad energy range covered by the combination of the 
NaI and BGO detectors allows a good measurement of the GRB prompt spectral parameters 
over a wide range. The broad energy range of GBM also permits investigation of possible deviations of the GRB spectra from standard empirical models (e.g. the Band function), which were in the past applied within a narrower energy domain. The instrument is described in detail by~\citet{Meegan:2009}. 
 
The very high temporal resolution of GBM is a major asset for the study of short events. The GBM onboard software incorporates burst triggering on time scales as short as 16 ms. All triggers generate Time-Tagged 
Event data (TTE) consisting of arrival times and the deposited energy for individual photons/events from each of the 14 detectors with a temporal resolution of 2~$\mu$s. A pre-burst ring buffer saves 512000 events before the trigger, which corresponds to~$\sim$30 seconds at background rates. TTE data are produced for 300 seconds after the trigger, thus all short bursts have full temporal coverage. The energy is digitized into 128 channels, pseudo-logarithmically 
spaced to provide channel widths less than the detector resolution up to 12 MeV, though TTE data are available at cruder spectral resolution up to 40 MeV.
 
Each of the GBM detectors has a fixed dead time per event of 2.6~$\mu$s, which is corrected in the present study. In addition, the spacecraft telemetry bus imposes a maximum total event rate (sum of all detectors) of 375k 
events per second. Neither of these limitations is significant for the bursts described here.

GBM detected about 70 short bursts between 2008, July 14 and 2009, October 1. Although its broad energy coverage, its good sensitivity above 1 MeV (due to the thickness of the BGO detectors), and its fine spectral resolution make GBM an ideal instrument to study GRB spectra, the instrument has a lower effective area than the {\it BATSE}/LADs and a higher background than the {\it Swift}/Burst Alert Telescope (BAT); as a  result most short bursts are not fluent enough for spectral analysis to be very constraining, particularly if one wishes to study spectral evolution over time. This low fluence problem is aggravated by
their typically higher $E_{\rm peak}$ values, which make both $E_{\rm peak}$ and the high-energy index of the Band function ($\beta$) more
difficult to measure for short bursts. In the following study, we have selected the GRBs detected with GBM from the start of the mission untill October 2009 with a fluence above $2 \times 10^{-6}$ ergs cm$^{-2}$ between 8 and 1000 keV. This fluence threshold assures the possibility of performing a time-resolved spectral analysis at the milli-second time-scale during the entire main emission of the burst. This selection criterion results in a sample of 3 bursts: GRBs~$090227$B,~$090228$~and~$090510$. Time-resolved analyses are still possible with other short GRBs detected with GBM, albeit in a coarser temporal and spectral resolution. The fourth brightest short GRB during this time period, GRB~{\it 081216}, with $1.4 \times 10^{-6}$ erg cm$^{-2}$, has 2.8 times less fluence than GRB~{\it 090510}, which limits the temporal resolution at which spectral analysis can be performed during the weaker portions of the GRB, requiring wider intervals that merge peaks and valleys of the light curve.

GRB~$090510$ triggered GBM and LAT on 2009 May 10 at 00:22:59.97 UT ~\citep{Guiriec:2009B} and 00:23:01.22 UT~\citep{Ohno:2009,Omodei:2009}, respectively. The burst was also detected with the {\it Swift}/BAT~\citep{Hoversten:2009} and located with the {\it Swift}/Ultra Violet Optical Telescope \citep[UVOT~--~][]{Kuin:2009} precisely enough to perform follow-up observations with optical ground-based telescopes. The Very Large Telescope/FOcal Reducer and low dispersion Spectrograph (VLT/FORS2) measured a spectroscopic redshift of 0.903$\pm$0.003~\citep{McBreen:2010,Rau:2009}.

GBM detected consecutively GRB~$090227$B~\citep{Guiriec:2009} and GRB~$090228$~\citep{VonKienlin:2009} on 2009 February 27 at 18:31:01.41 UT 
and on 2009 February 28 at 04:53:20.91 UT, respectively. These bright events were also seen by several other instruments: Konus-Wind, Suzaku-WAM, INTEGRAL SPI-ACS,and MESSENGER for GRB~$090227$B; Konus-Wind, Mars Odyssey, MESSENGER and Agile-MCAL for GRB~$090228$. This allowed reconstruction of the arrival direction of their photons by the Inter--Planetary Network (IPN) and provided an error box for the position of each GRB (Hurley et al.: 3$^{rd}$ IPN Catalog in preparation). These locations typically 
achieve more precision than is possible with GBM alone. The best locations for all three GRBs are reported in Table~\ref{table:locations}. The set of the brightest GBM detectors with an angle to each source below 80{\degree}~and a source view not blocked by the spacecraft's components (solar panels, radiators, LAT, etc.) are reported in Table~\ref{table:detectors}.
  
\begin{table}[h!]
\begin{center}
\caption{\label{table:locations} Best locations obtained for GRBs~$090227$B, $090228$ and $090510$.}
\begin{tabular}{|l|c|c|c|c|}
\hline
GRB & Instrument(s) & RA (\degree) & DEC (\degree) & ERR \\
\hline
\hline
GRB~$090227$B & IPN        & 15.6      & 26.4      & 0.6$^a$ \\
GRB~$090228$  & IPN        & 98.3      & -28.4     & 0.0017$^a$\\
GRB~$090510$  & {\it Swift}/UVOT & 333.6 & -26.6 & 1.5$^b$ \\ 
\hline
            & \multicolumn{4}{|c|}{$^a$ in square degrees; $^b$ in arc seconds}\\
\hline
\end{tabular}
\end{center}
\end{table}

\begin{table}[h!]
\begin{center}
\caption{\label{table:detectors}Detectors used for the spectral analysis in each GRB. We used the brightest detectors with an angle to the source below 80{\degree}~for the NaI detectors and the closest detector to the source for the BGOs.}
\vspace{0.3cm}
\begin{tabular}{|l|c|c|}
\hline
GRB & NaI detectors & BGO detectors \\
\hline
\hline
GRB~$090227$B  & n0, n1, n2, n5     & b0     \\
GRB~$090228$   & n0, n1, n2, n3, n5 & b0     \\
GRB~$090510$   & n3, n6, n7, n8, n9 & b0, b1 \\
\hline
\end{tabular}
\end{center}
\end{table}  
  
\section{Observations: I. Temporal analysis}
\label{section:observation}

\subsection{Light-Curves and Time lags}
\label{subsection:lightcurve}

\begin{figure}[h!]
\begin{center}
\includegraphics*[viewport=0 10 520 293,width=0.5\textwidth]{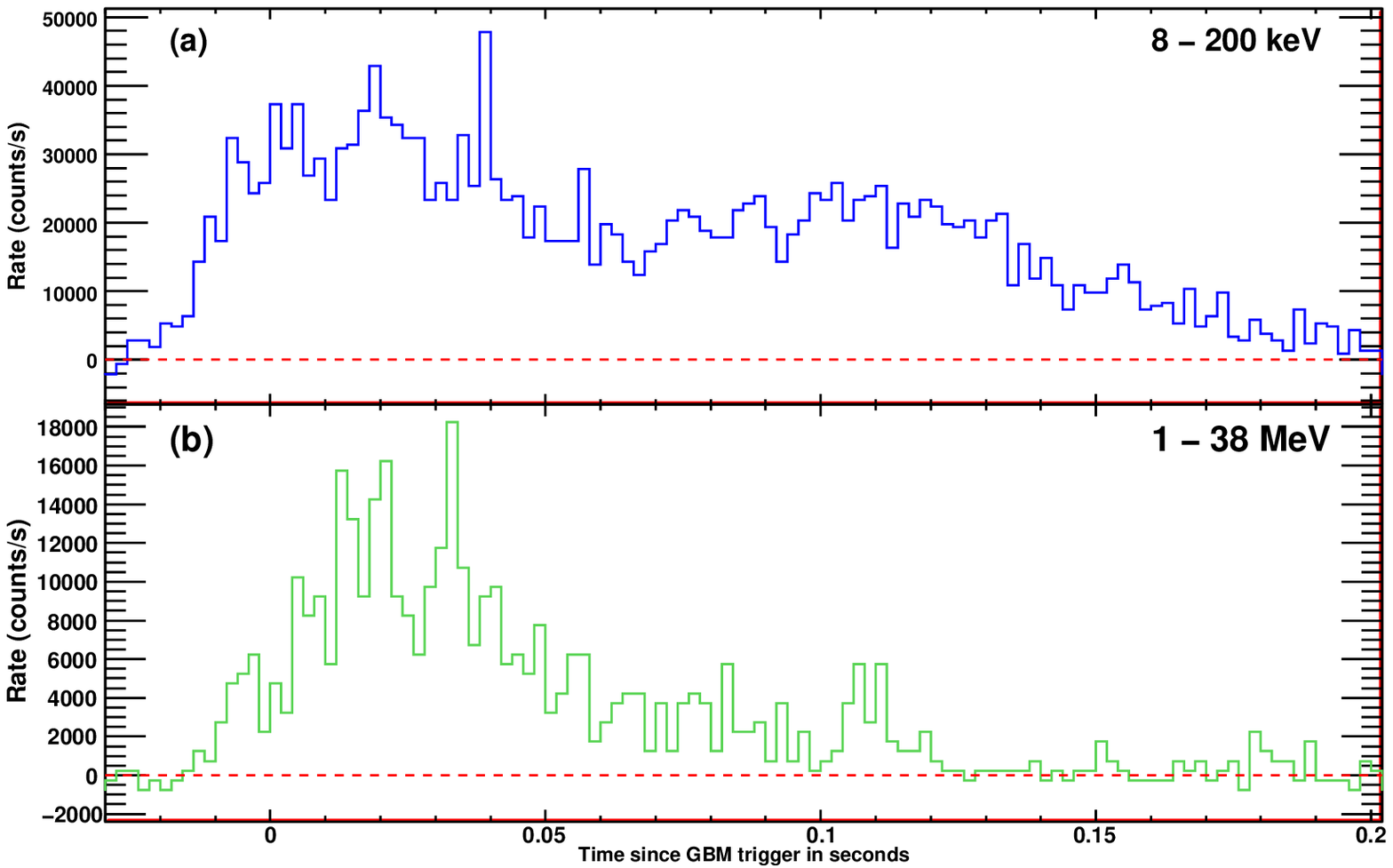}\\
\vspace{-0.24cm}
\includegraphics*[viewport=0 0 520 293,width=0.5\textwidth]{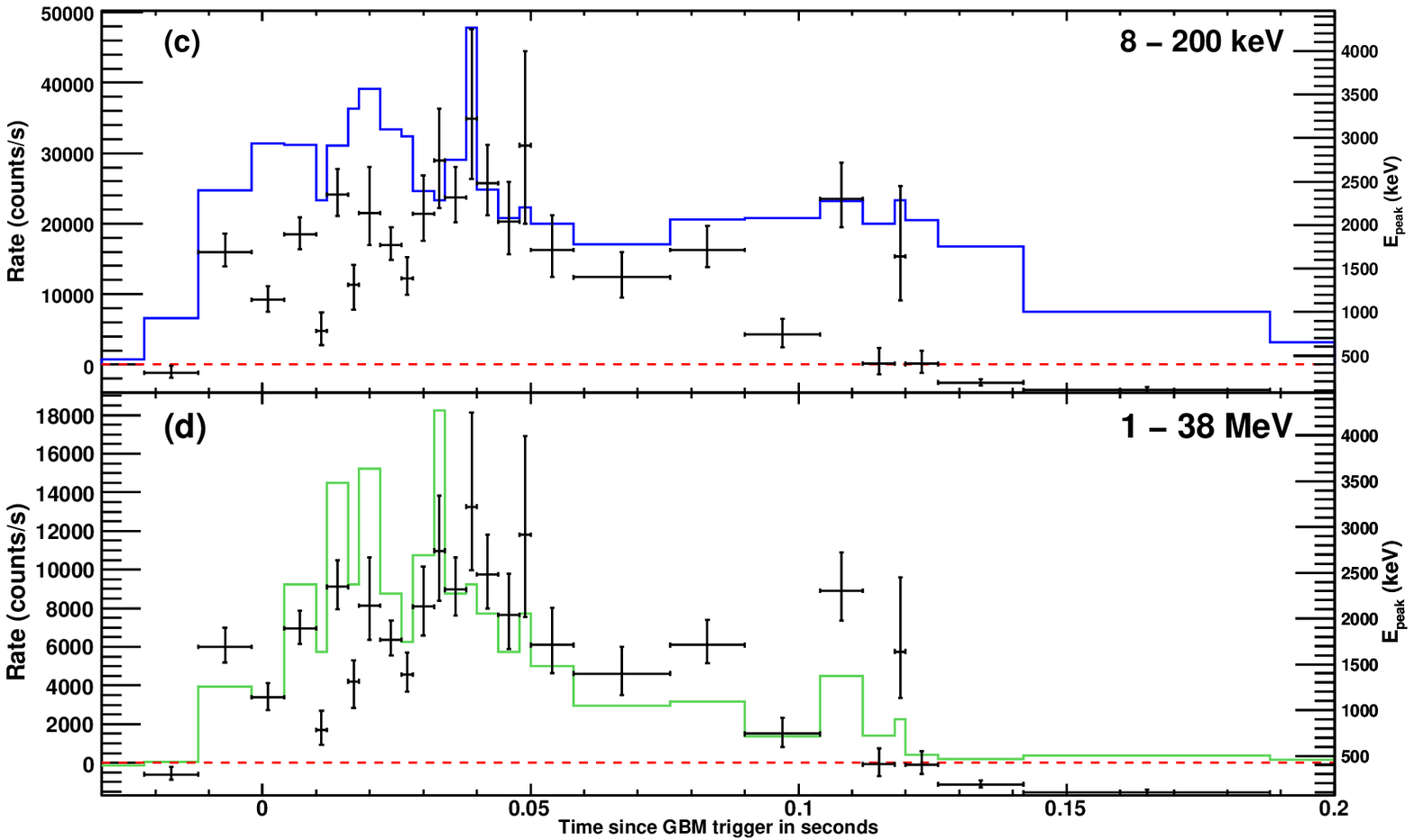}
\caption{Light curves of GRB~$090227$B in two energy bands ({\it Panel a:} 8 keV to 200 keV, NaI detectors) and ({\it Panel b:} 1 MeV to 38 MeV, BGO detectors) with 2 ms time resolution. The count rates are background substracted. {\it Two bottom panels:} The same light-curves with variable time bins (histograms), optimized for time-resolved spectroscopy. The Band function peak energy, $E_{\rm peak}$, is plotted over the lightcurve for each time interval.\label{fig:GRB090227B_EpeakOverLC}}
\end{center}
\end{figure}

\begin{figure}[h!]
\begin{center}
\includegraphics*[viewport=0 10 520 293,width=0.5\textwidth]{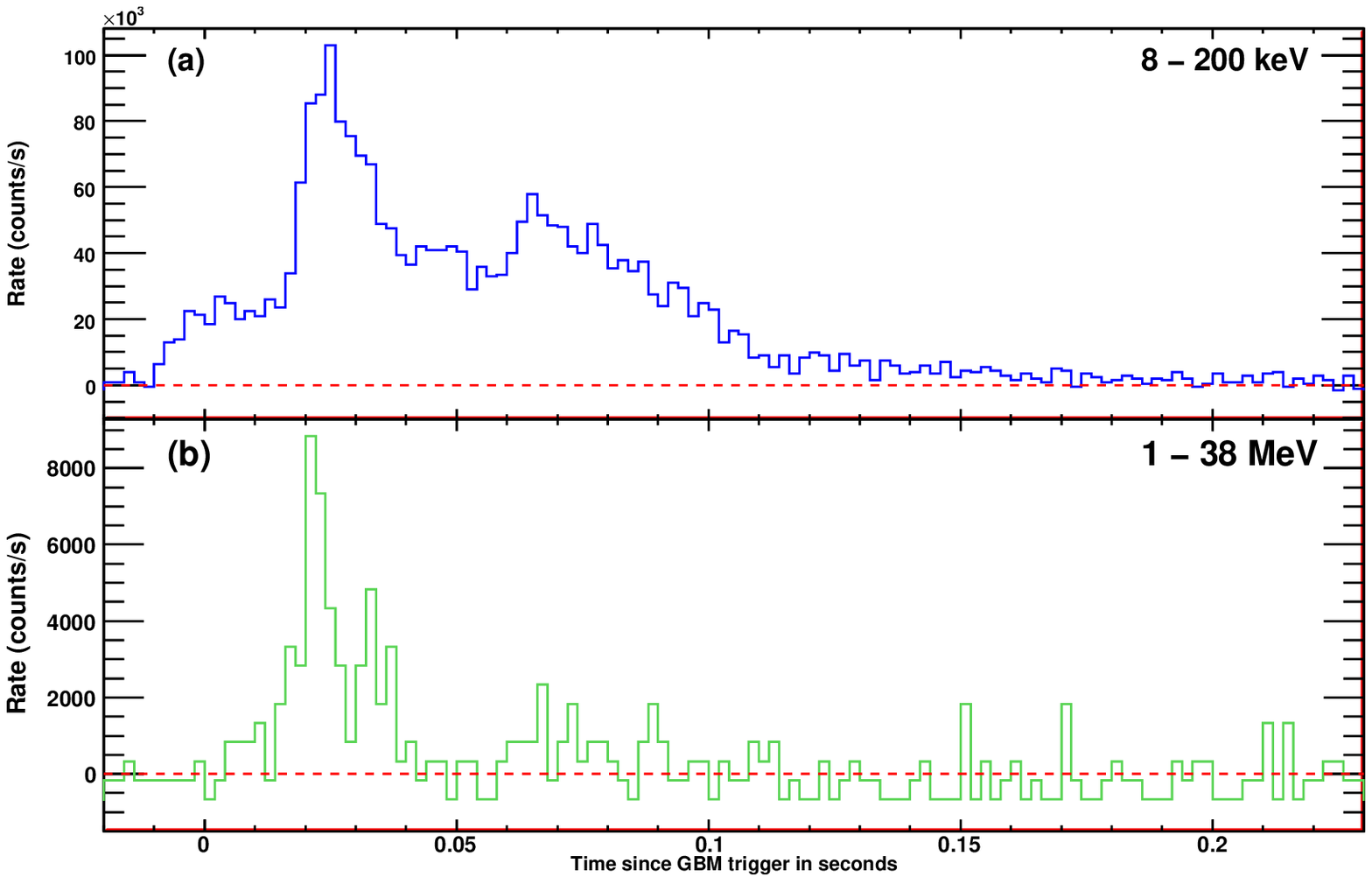}\\
\vspace{-0.24cm}
\includegraphics*[viewport=0 0 520 293,width=0.5\textwidth]{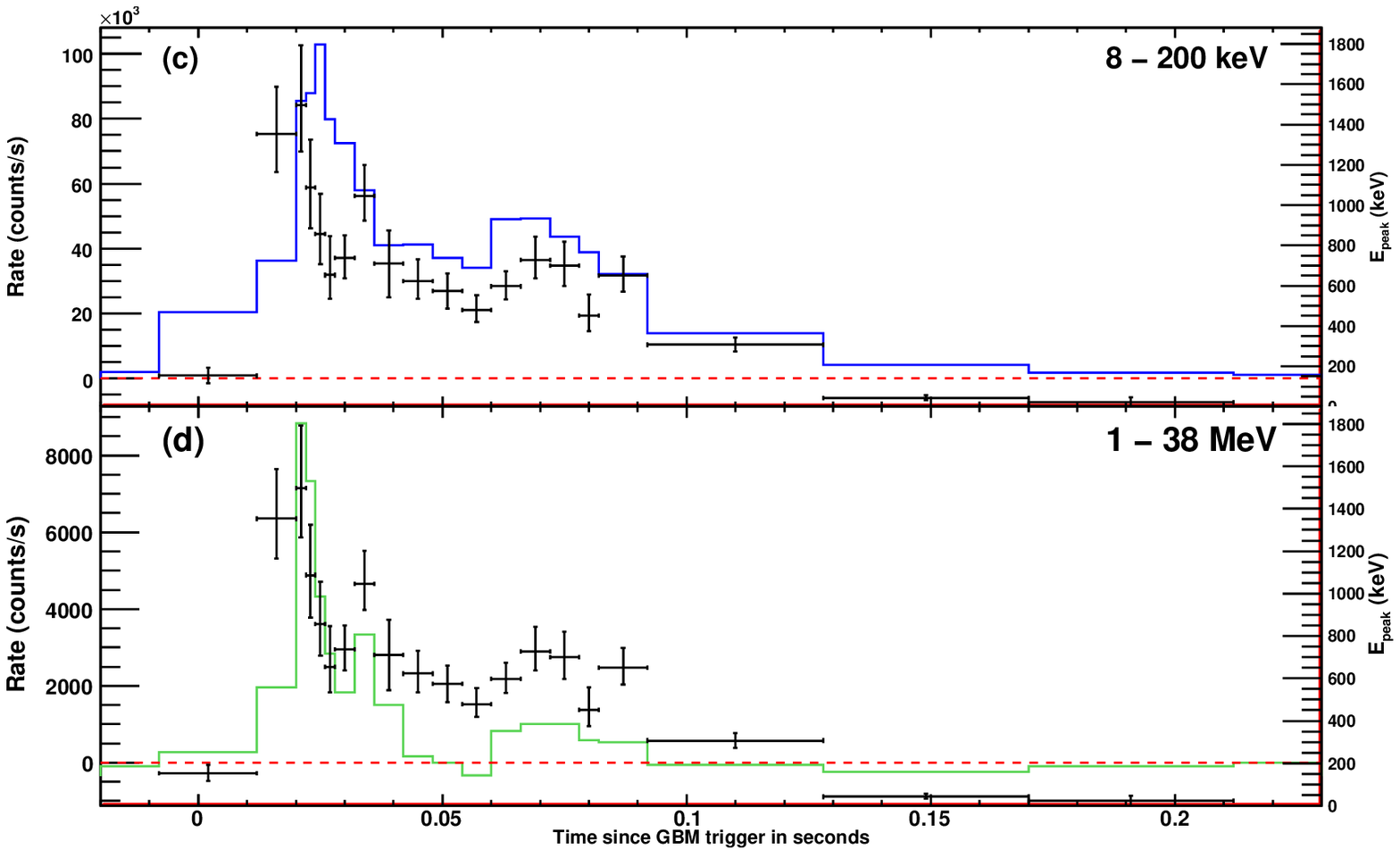}
\caption{Light curves of GRB~$090228$ in two energy bands ({\it Panel a:} 8 keV to 200 keV, NaI detectors) and ({\it Panel b:} 1 MeV to 38 MeV, BGO detectors) with 2 ms time resolution. The count rates are background substracted. {\it Two bottom panels:} The same light-curves with variable time bins (histograms), optimized for time-resolved spectroscopy. The Band function peak energy, $E_{\rm peak}$, is plotted over the lightcurve for each time interval. \label{fig:GRB090228_EpeakOverLC}}
\end{center}
\end{figure}

\begin{figure}[h!]
\begin{center}
\includegraphics*[viewport=0 10 520 293,width=0.5\textwidth]{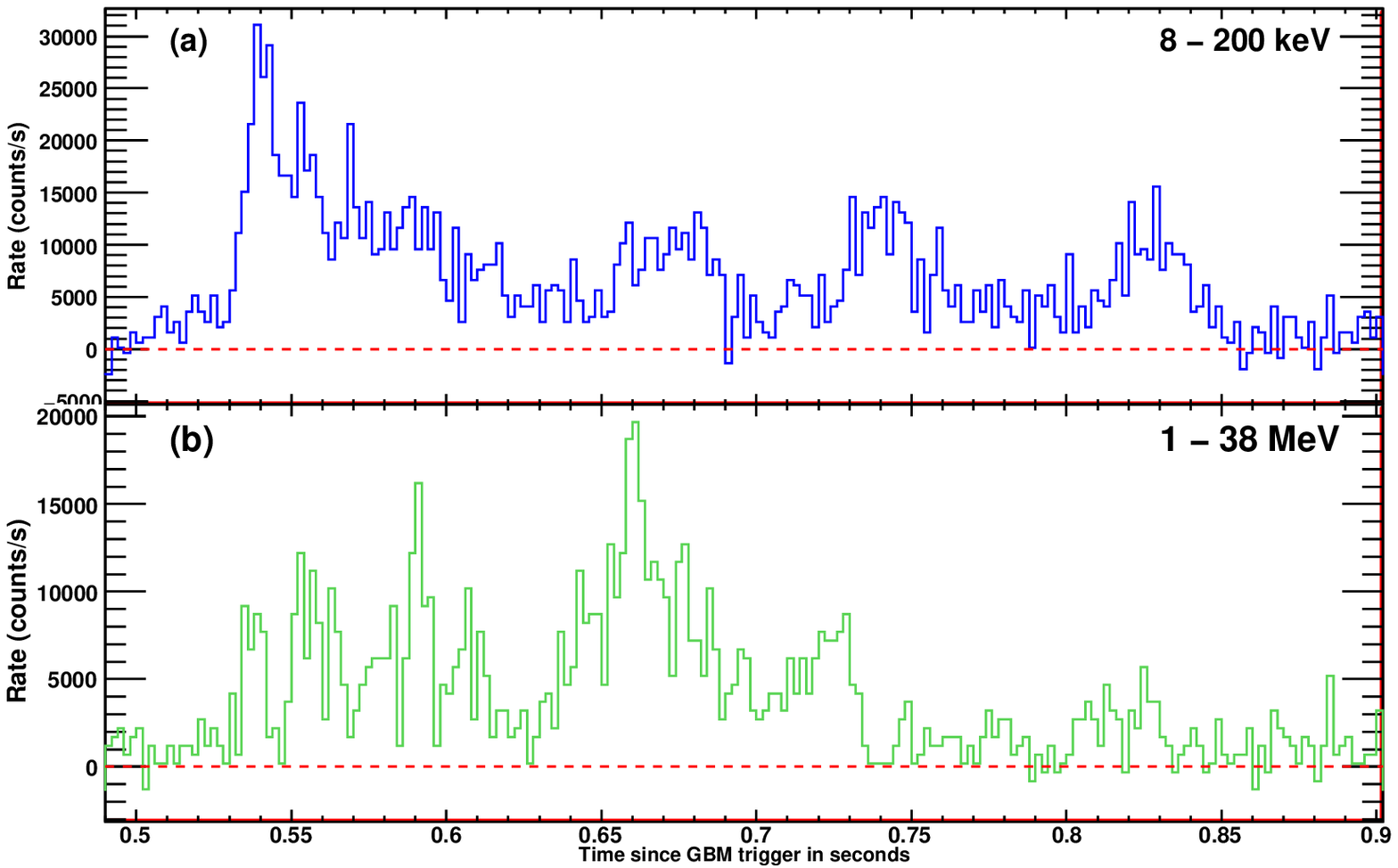}\\
\vspace{-0.24cm}
\includegraphics*[viewport=0 0 520 293,width=0.5\textwidth]{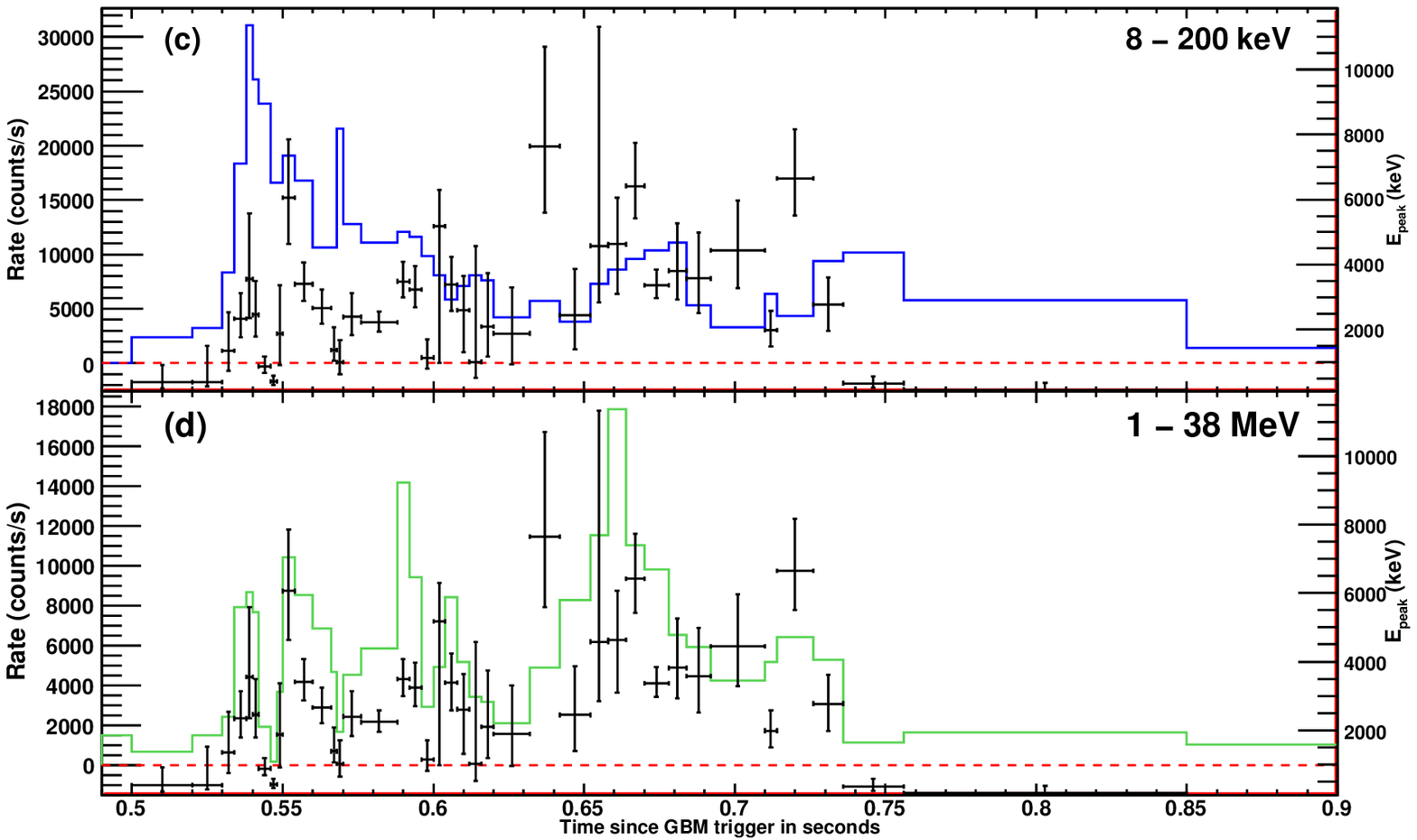}
\caption{Light curves of GRB~$090510$ in two energy bands ({\it Panel a:} 8 keV to 200 keV, NaI detectors) and ({\it Panel b:} 1 MeV to 38 MeV, BGO detectors) with 2 ms time resolution. The count rates are background substracted. {\it Two bottom panels:} The same light-curves with variable time bins (histograms), optimized for time-resolved spectroscopy. The Band function peak energy, $E_{\rm peak}$, is plotted over the lightcurve for each time interval. \label{fig:GRB090510_EpeakOverLC}}
\end{center}
\end{figure}

The high temporal resolution (2 ms) light-curves of all three GRBs are shown in figures~\ref{fig:GRB090227B_EpeakOverLC}~to~\ref{fig:GRB090510_EpeakOverLC} in two energy ranges, $8-200$ keV ({\it panels a}) and 1--38 MeV ({\it panels b}), respectively. GRB~$090510$ (Figure~\ref{fig:GRB090510_EpeakOverLC}) triggered GBM at $T_0$ on a very soft and weak pulse, not shown in the plot, which was followed by an interval of about 0.5 s without significant emission. Figure~\ref{fig:GRB090510_EpeakOverLC} shows the main emission from this burst starting at $T_0+0.5$ s.

For all three GRBs, the higher energy light curves exhibit much sharper structures than the lower energy curves, where the emission is smoother. GRBs~$090227$B and $090510$ show very fast variability with multiple spikes in all energy bands; GRB~$090228$ is much simpler with only two main peaks. Overall, the signal rises much faster than it decays. A visual comparison of the high and low energy light curves reveals a trend of soft-hard-soft evolution in all events, which is also supported by our spectral fits as discussed in section~\ref{Epeak_evolution_and_distribution}. The most dramatic evolution is demonstrated in the data of GRB~$090510$~(Figure~\ref{fig:GRB090510_EpeakOverLC}).  


\begin{figure}
\begin{center}
\includegraphics[width=0.48\textwidth]{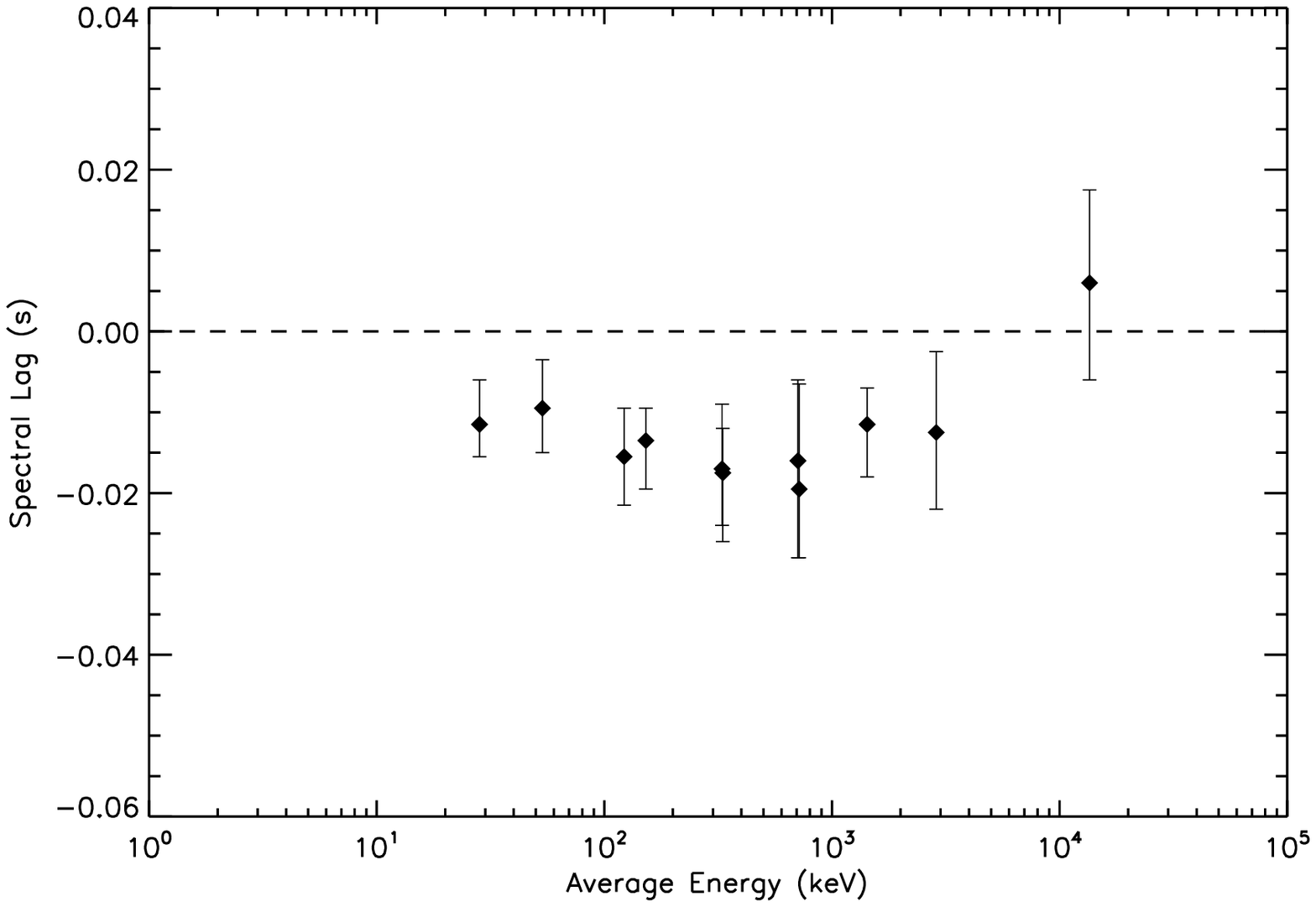}\\
\includegraphics[width=0.48\textwidth]{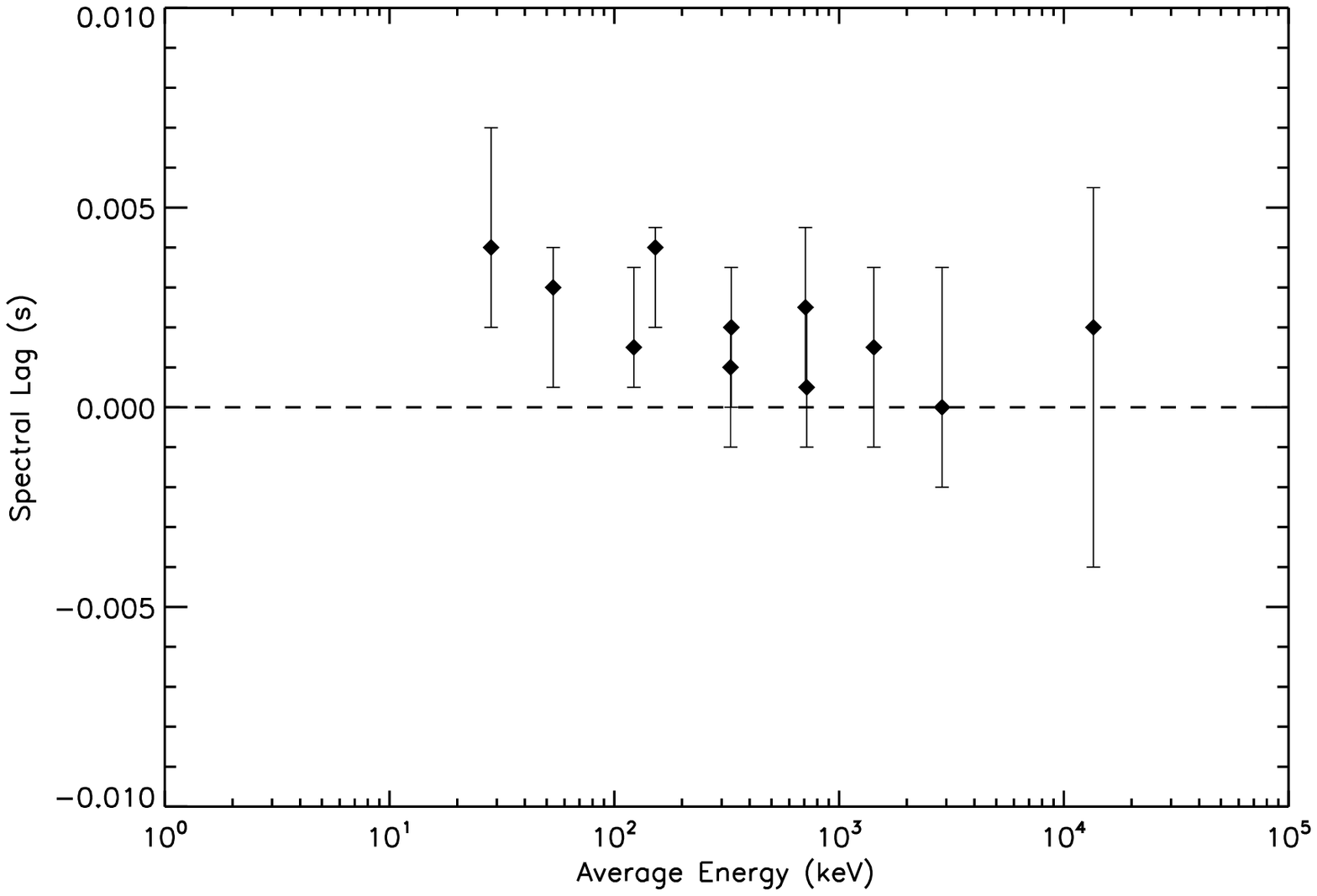}\\
\includegraphics[width=0.48\textwidth]{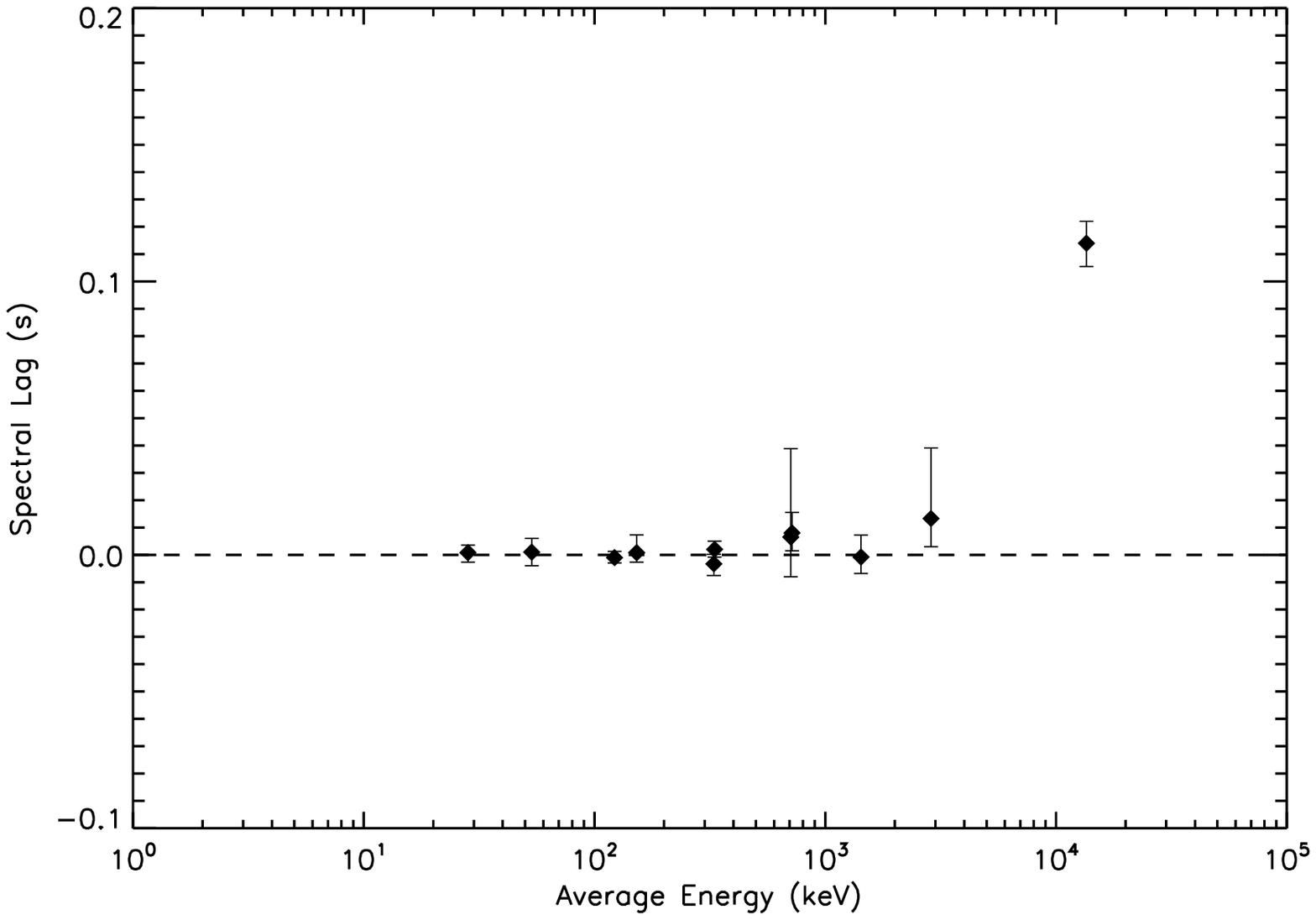}
\end{center}
\caption{Spectral lags for GRB~$090227$B (top panel), GRB~$090228$ (middle panel) and GRB~$090510$ (bottom panel). All lags were computed with respect to the 8--20 keV NaI energy band (see text for details). The temporal resolution was 2 ms for GRBs~$090227$B and ~$090228$, and 1 ms for GRB~$090510$.
\label{figure:lags}}
\end{figure}

We have estimated the time averaged spectral lags for each of the events using a cross-correlation technique \citep{Cheng:1995,Band:1997}. We use the Pearson cross-correlation function to estimate the degree of correlation of two independent time series. For each detector type (NaI and BGO), we generated the GRB light curves using the TTE data divided into 6 energy bands: 4--20 keV, 20--41 keV, 41--70 keV, 70--212 keV, 212-510 keV, and 510--985 keV (NaIs); 109--212 keV, 212--517 keV, 517--996 keV, 996--2048 keV, 2.0--4.0 MeV, and 4--45.5 MeV (BGOs). We then measured the lags between the first NaI energy band and each of the others. To evaluate the errors on the lags, we performed the same analysis with simulated light-curves generated from the real light curves by adding Poisson fluctuations based on the count rate in each time bin.  The lag errors are estimated from the distribution of 100 such trials. The results are presented in Figure~\ref{figure:lags}. Negative (positive) numbers indicate that the higher energy photons lead (lag) the low energy photons.

In general, short GRBs are expected to exhibit small or no spectral lags below 1--2 MeV~\citep{Norris:2006}. Figure~\ref{figure:lags} shows that for GRB~$090227$B there is a constant lag of $\sim15$ ms (1-2$\sigma$ level) between the higher energy (30$-$3000 keV) photons and those at $\sim$8--20 keV . 
For GRB~$090228$, although the lag values are all positive (indicating a lag of higher energies with respect to low energies) the errors are too large to allow a significant measurement. Finally, when the burst has sufficient counts in the higher energy bands ($>$ ~few MeV), for GRB~$090510$, we measure a positive lag between 8 keV to 3 MeV and the highest energy band ($>3$ MeV). This result is also supported by the higher energy LAT data,  as already published in the supplementary information of~\cite{Abdo:2009}. The energy range 8--20 keV is outside that used in~\citet{Norris:2000}. The canonical energy channels used to provide a discriminant between short and long bursts show no evidence for lags in any of the three bursts, consistent with the~\citet{Norris:2006} observation that short bursts show small/zero-lags.

The average lags measured above indicate an overall good alignment of the light-curves from 20 keV to several MeV. However, local time lags and leads could still be present between individual pulses, as previously observed for a number of long {\it BATSE} GRBs~\citep{Ryde:2005,Hakkila:2008}. In particular, we note that, as also seen in the two upper panels (a and b) of Figure~\ref{fig:GRB090228_EpeakOverLC},  in GRB~$090228$ the peak of the brightest pulse in 8--200 keV lags the peak in the $>$1 MeV band. Such local delays cancel out when integrated over the entire light curve, resulting in a small or zero lag.

\subsection{Durations}
\label{section:timing}

We have performed a detailed temporal analysis of the three selected GRBs, applying the technique described in~\citet{Koshut:1996} to 
determine the $T_{90}$ and $T_{50}$ durations of all three events in the 50--300 keV energy range using the TTE data 
type (see section~\ref{section:GBM}). $T_{90}$ ($T_{50}$) is the time between accumulating 
5\% and 95\% (25\% and 75\%) of the counts associated with the GRB~\citep{Kouveliotou:1993}. We describe below only the results of this analysis for GRBs~$090227$B and~$090228$; the duration analysis of GRB~$090510$ has been 
described in the Supplementary Material of the paper by~\citet{Abdo:2009}. For both events we used in our analysis the combined light curves of the three NaI detectors 
with the smallest zenith angles to the source. For GRB~$090227$B these are NaI 0, 1, and 2 with angles ranging from 19{\degree} to 
52{\degree}; for GRB~$090228$ these are NaI 0, 1, and 3, ranging from 8{\degree} to 42{\degree}. Table \ref{table:durations} contains 
$T_{90}$~and~$T_{50}$~results for all three events. 

\begin{table}
\caption{\label{table:durations} $T_{90}$ and $T_{50}$ durations.}
\begin{center}
\begin{tabular}{|l|c|c|}
\hline
Burst & $T_{90}$ (s) & $T_{50}$ (s) \\
\hline
\hline
GRB~$090227$B  & 2.38   & 0.09 \\
GRB~$090228$   & 0.34   & 0.05 \\
GRB~$090510$   & 2.1    & 0.2  \\
\hline
\end{tabular}
\end{center}
\end{table}

\begin{figure}[h!]
\begin{center}
\includegraphics[height=0.5\textwidth,angle=-90]{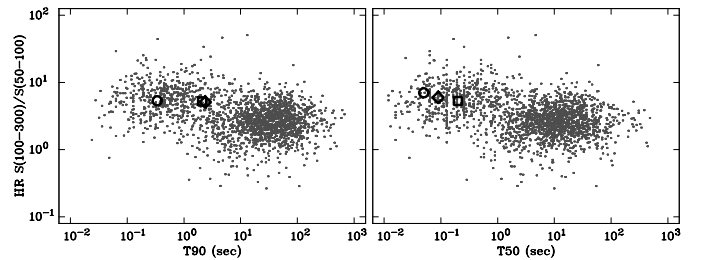}
\end{center}
\caption{Hardness-duration diagrams ({\it Left Panel:} $T_{90}$; {\it Right Panel:}  $T_{50}$) for 1973 events of the 4B {\it BATSE} catalog (black dots). The three GBM short GRBs are overplotted: GRB~$090227$B (diamond), GRB~$090228$ (circle) and GRB~$090510$ (square). 
\label{figure:hardness-duration}}
\end{figure}

Further, we computed the fluence hardness ratios (HR) of all events over the $T_{90}$ and $T_{50}$ durations by fitting their spectra 
with the spectral functions described in section~\ref{section:time-integrated-spectral-analysis}. These ratios were computed between the 
$100-300$ keV and the $50-100$ keV energy bands from deconvolved spectra for direct comparisons with the GRBs in the {\it BATSE} 4B catalog~\citep{Paciesas:1999}. 
Figure~\ref{figure:hardness-duration} displays the HR values of the three GBM events superposed on the HR diagram of 1973 events with 
well determined durations, HRs and fluences from the 4B catalog. We notice that for all three events the HRs are well within those for the 
{\it BATSE} short GRB population, while the durations have a wider spread. For two of the events, the $T_{90}$ values are in the 
overlapping area between long and short GRBs, while their $T_{50}$ values are well within the short GRB domain. This spread demonstrates that 
the former duration measure is very sensitive to 5\% background variations, while the latter is more robust. Figure~\ref{figure:hardness-duration} also demonstrates that all three events are within the short GRB range, albeit, two of them at the long tail of the 
distribution.

We note that a third class of intermediate GRBs has been suggested. \citet{Hakkila:2004}, however, questioned the interpretation as a physical class, finding, instead, that the putative class is caused by instrumental selection effects. Following the classification scheme of~\citet{Horvath:2006}, all three GRBs fall in the short class of GRBs.

%
%
%

\section{Observations: II. Spectral Analysis}
\label{section:spectral-analysis}

We performed simultaneous fits of the TTE data (see section~\ref{section:GBM}) of NaI and BGO detectors selected for each of the three short GRBs (see Table~\ref{table:detectors}). We fit the data with various spectral functions (see next section) and used the Castor C-stat statistic to select the best fit in each case. The Castor C-stat differs from the Poisson likelihood statistic by an offset which is a constant for a particular dataset; its distribution is similar to a $\chi^{2}$ distribution for high statistical-content regimes. We chose Castor C-stat because of its robustness in the low counts regime, which is the case for these 3 GRBs, especially for the time-resolved fits of the data. All spectral analyses were performed using RMFIT version 3.1 and GBM Response Matrices version 1.8. The detector responses were calculated for the best available location of the GRBs, either from Swift or the IPN (see 
Table~\ref{table:locations}). Backgrounds were estimated by fitting low-order polynomials to time intervals before and after each event.

\subsection{Time-Integrated Spectral Analysis}
\label{section:time-integrated-spectral-analysis}

Table~\ref{table:integrated_spectra} summarizes the results of our time-integrated analysis. For each of the three GRBs we selected intervals longer than their $T_{50}$ durations, as indicated in the Table caption. As in most of the previous spectral analyses of short GRBs, we used either the ``comptonized'' model (Compt), or the Band GRB function \citep{Band:1993}. The Compt model consists of a power law times an exponential cutoff, and has 3 parameters. In our study, this cutoff is expressed as $E_{\rm peak}$, namely the location of the peak of 
the ${\nu}F_{\nu}$ spectrum~\citep{Gehrels:1997}. We also searched for deviations of the spectra from a single component by adding a power law (PL) to 
each of these models. 

\subsubsection{Identification of an additional component}

For all three GRBs, the combination of Band+PL provides a significantly better fit than the Band model alone. Similarly, for all three 
GRBs, the combination of Compt+PL is significantly better than the Compt model alone. Table~\ref{table:cstat_improvement} shows the 
improvement in Castor C-stat for the more complex models and the corresponding significance of this improvement. 
All significance values reported herein are for comparing nested models, for which the probability value can be calculated from the 
difference in Castor C-stat from the $\chi^{2}$ distribution for the change in the number of model parameters.
Testing Compt+PL and Band+PL, one finds that, either the two models are tied (GRBs~$090228$ and~$090510$) or the improvement to the fitting 
statistic is not large enough to justify the more complex Band function (GRB~$090227$B, $\Delta$ Castor C-stat = 3, $P=0.08$).

\begin{table}[h!]
\caption{Improvement in Castor C-stat with addition of Power-Law component\label{table:cstat_improvement}}
\begin{center}
{\tiny
\begin{tabular}{|l|l|c|l|}
\hline
\multicolumn{1}{|c|}{Name} & \multicolumn{1}{|c|}{Model} & \multicolumn{1}{|c|}{$\Delta$C-Stat} & \multicolumn{1}{|c|}{Significance} \\
\hline
GRB~$090227$B & Compt+PL over Compt    & 17 & $2.0 \times 10^{-4}$ \\
            & Band+PL over Band      & 13 & $1.5 \times 10^{-3}$ \\
            & Band+PL over Compt+PL  &  3 & $ 0.083$ \\
\hline
GRB~$090228$ & Compt+PL over Compt    & 18 & $1.2 \times 10^{-4}$ \\
            & Band+PL over Band      & 18 & $1.2 \times 10^{-4}$ \\
            & Band+PL over Compt+PL  &  0 & $ 1.0$ \\
\hline
GRB~$090510$ & Compt+PL over Compt    & 25 & $3.7 \times 10^{-6}$ \\
            & Band+PL over Band      & 14 & $9.1 \times 10^{-4}$ \\
            & Band+PL over Compt+PL  &  0  & $ 1.0$ \\
\hline
\end{tabular}
}
\end{center}
\end{table}

\begin{figure}
\includegraphics[totalheight=0.17\textheight, viewport=8 0 540 730,clip,angle=90]{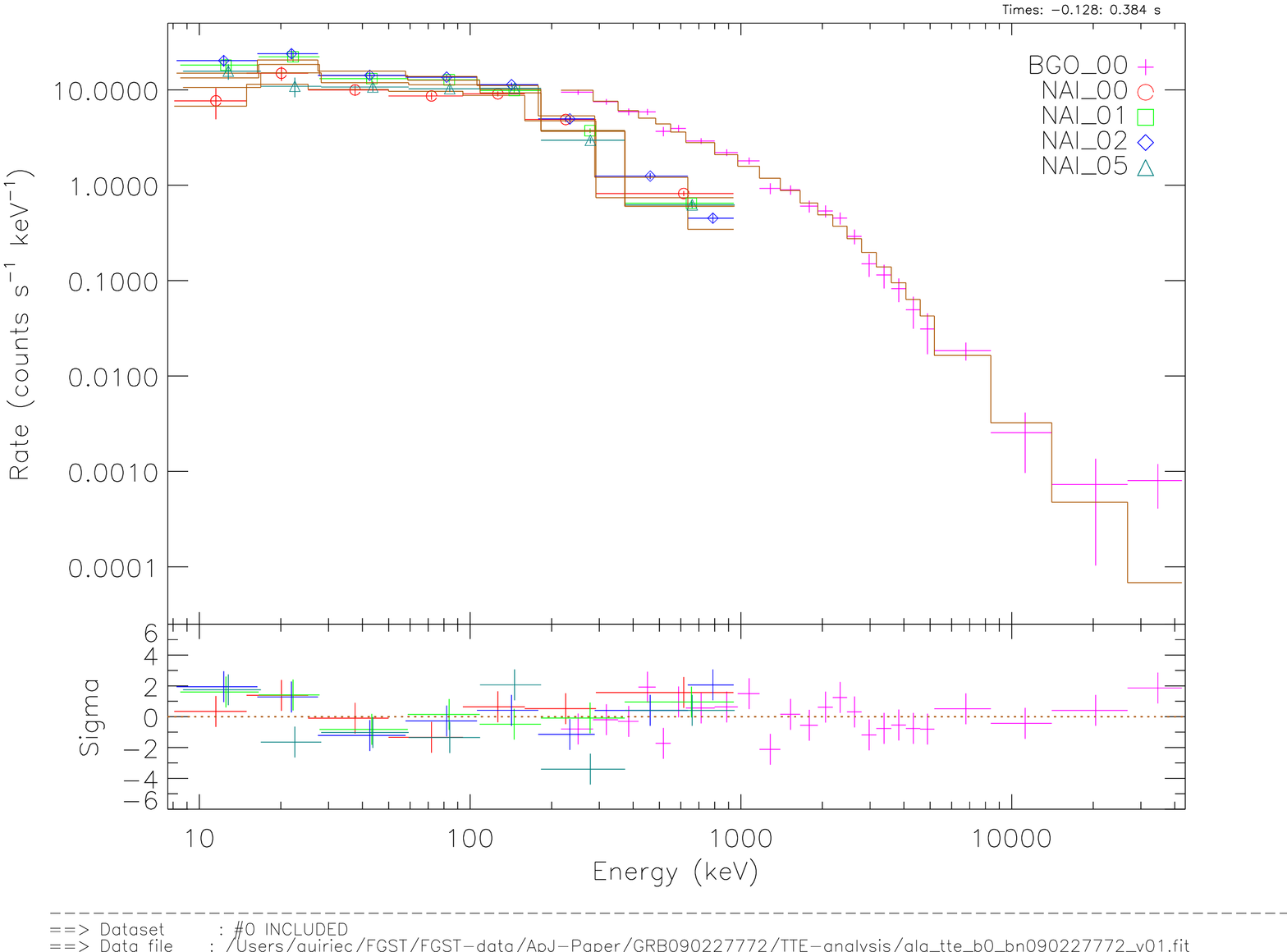}
\includegraphics[totalheight=0.17\textheight, viewport=8 0 540 730,clip,angle=90]{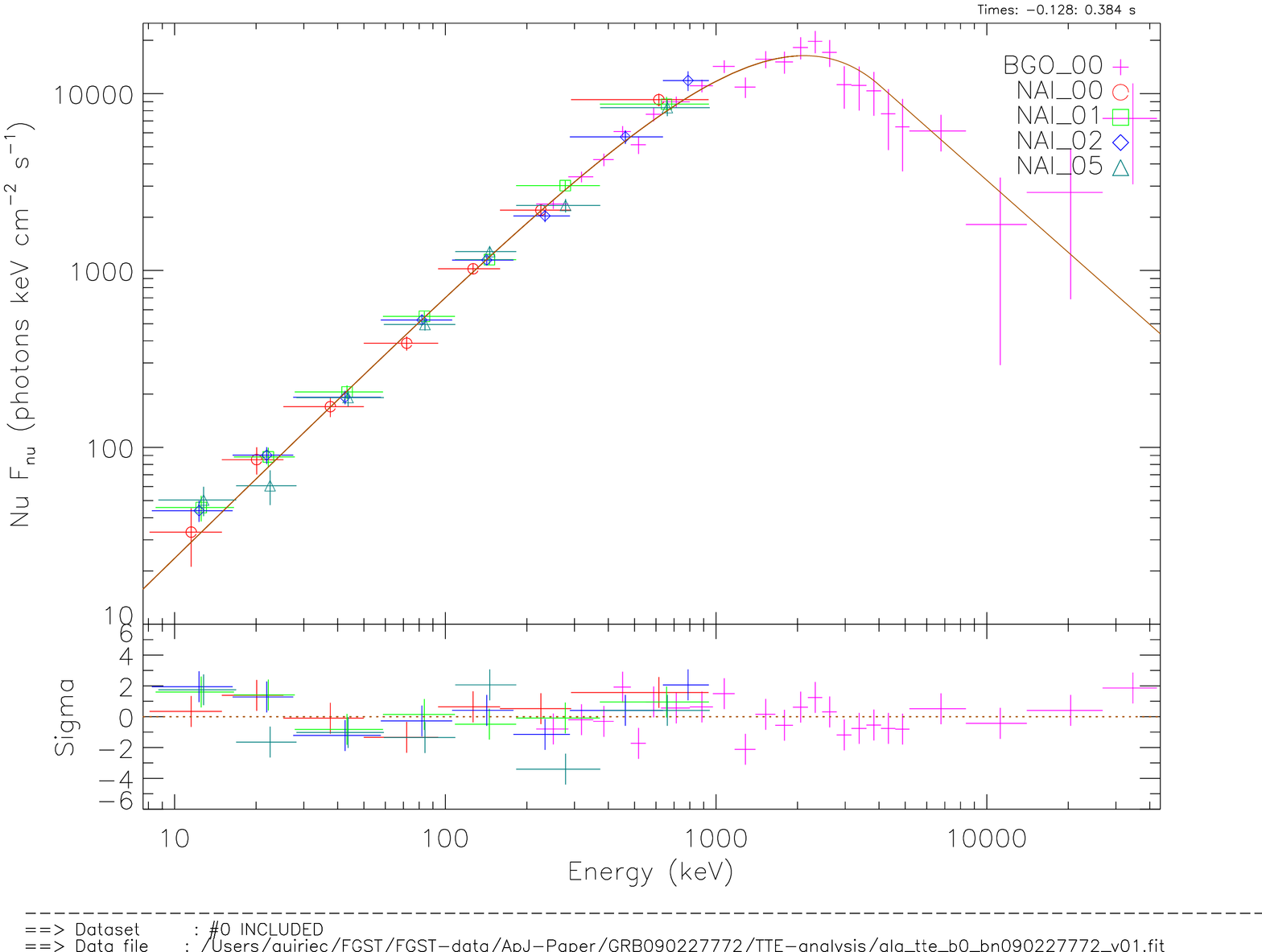}

\includegraphics[totalheight=0.17\textheight, viewport=8 0 540 730,clip,angle=90]{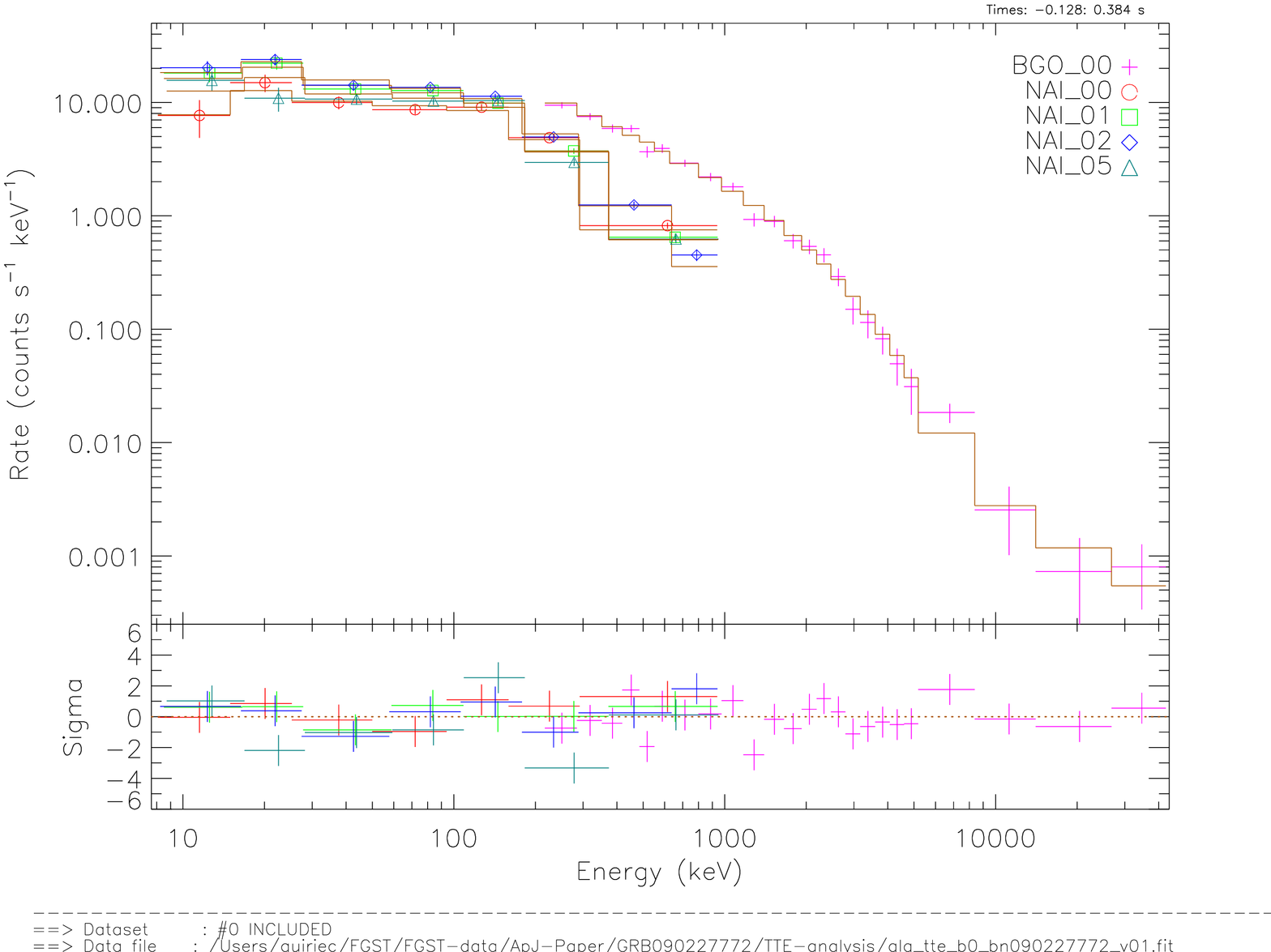}
\includegraphics[totalheight=0.17\textheight, viewport=8 0 540 730,clip,angle=90]{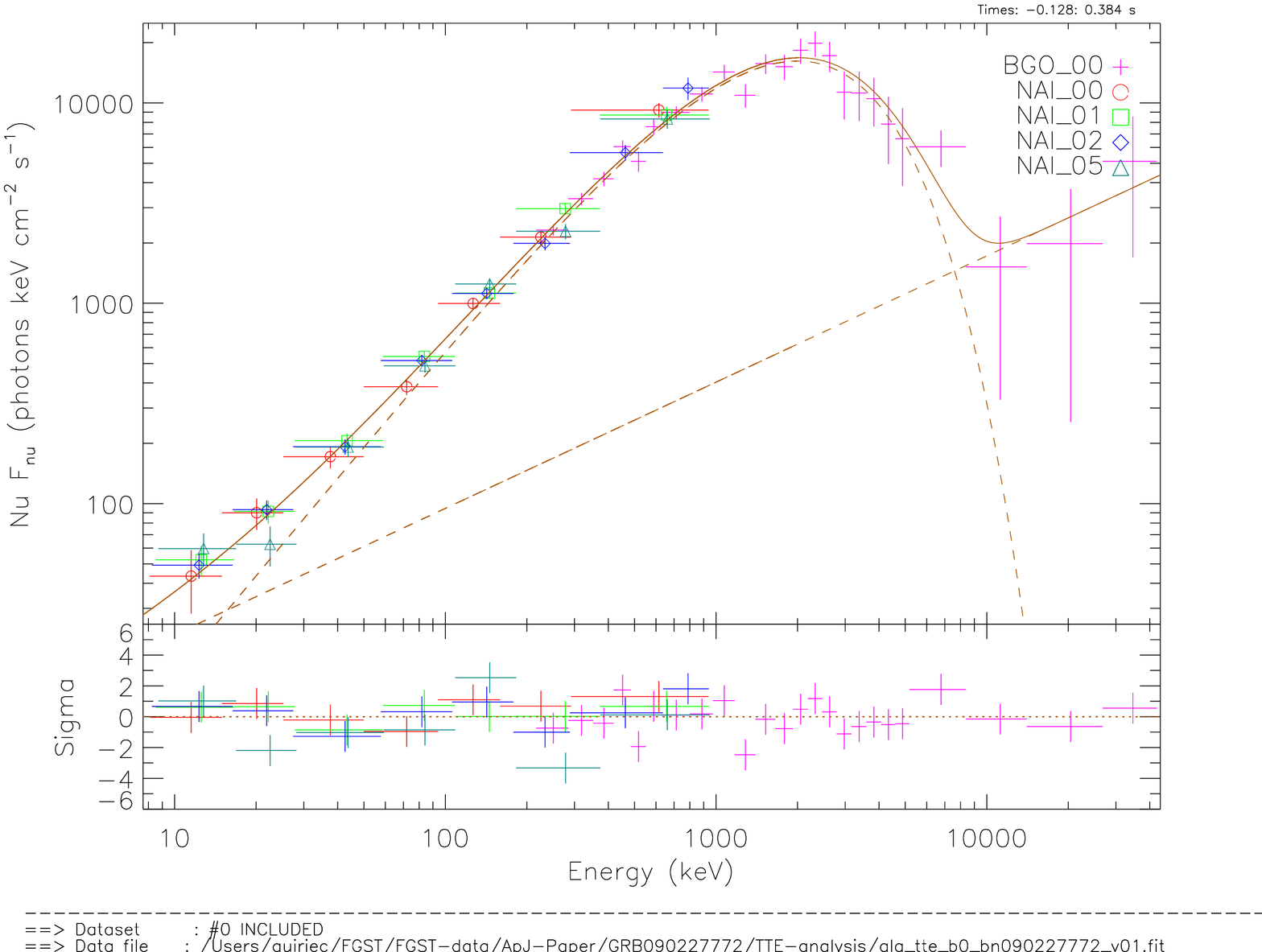}

\includegraphics[totalheight=0.17\textheight, viewport=8 0 540 730,clip,angle=90]{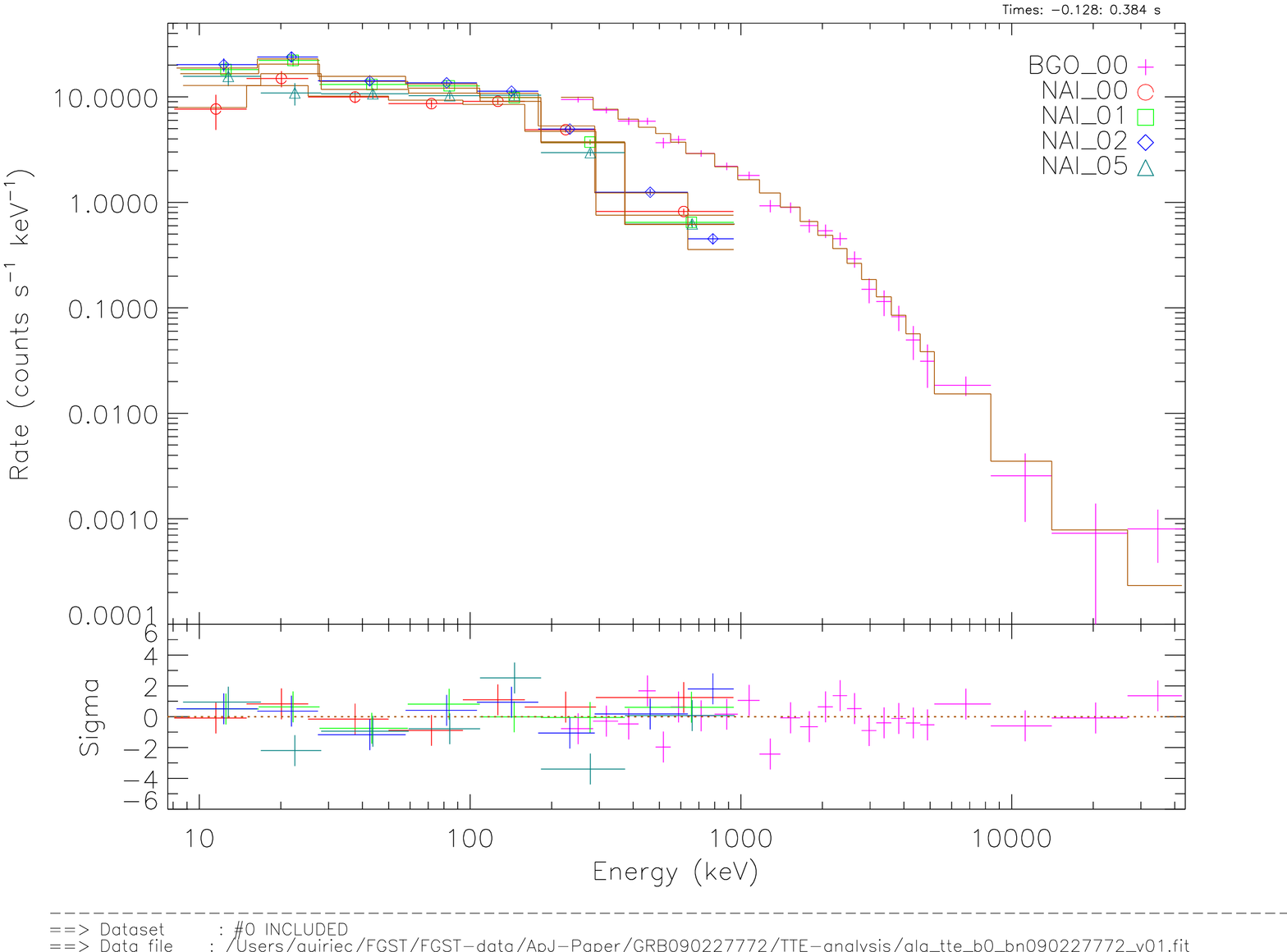}
\includegraphics[totalheight=0.17\textheight, viewport=8 0 540 730,clip,angle=90]{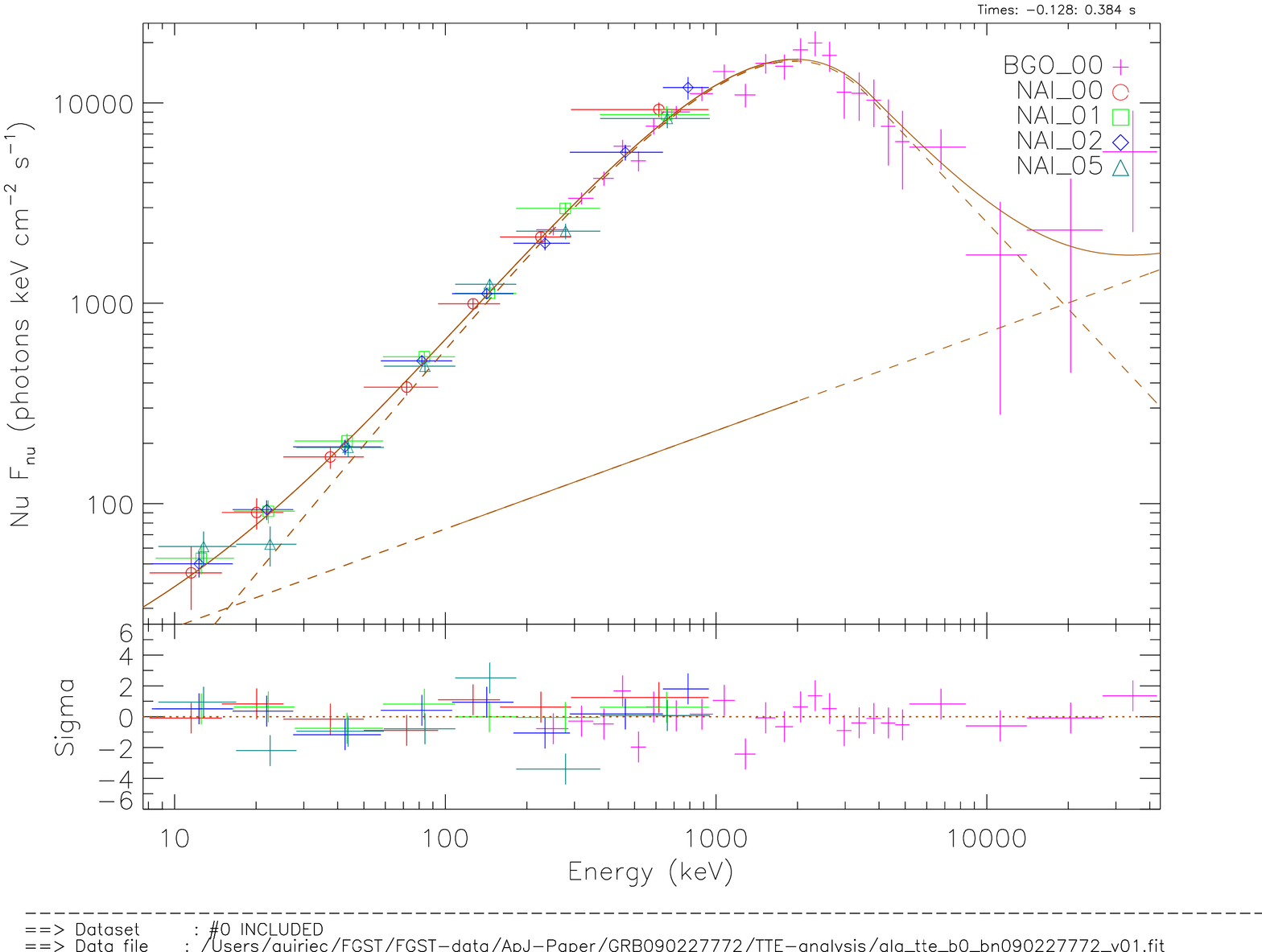}
\caption{Integrated count spectra ({\it left column}) and ${\nu}F_{\nu}$ spectra ({\it right column}) for GRB~$090227$B with Band ($1^{st}$ row), Compt+PL ($2^{nd}$ row) and Band+PL ($3^{rd}$ row) spectral fits.\label{fig:GRB090227B_integrated_spectra}}
\end{figure}

\begin{figure}
\includegraphics[totalheight=0.17\textheight, viewport=8 0 540 730,clip,angle=90]{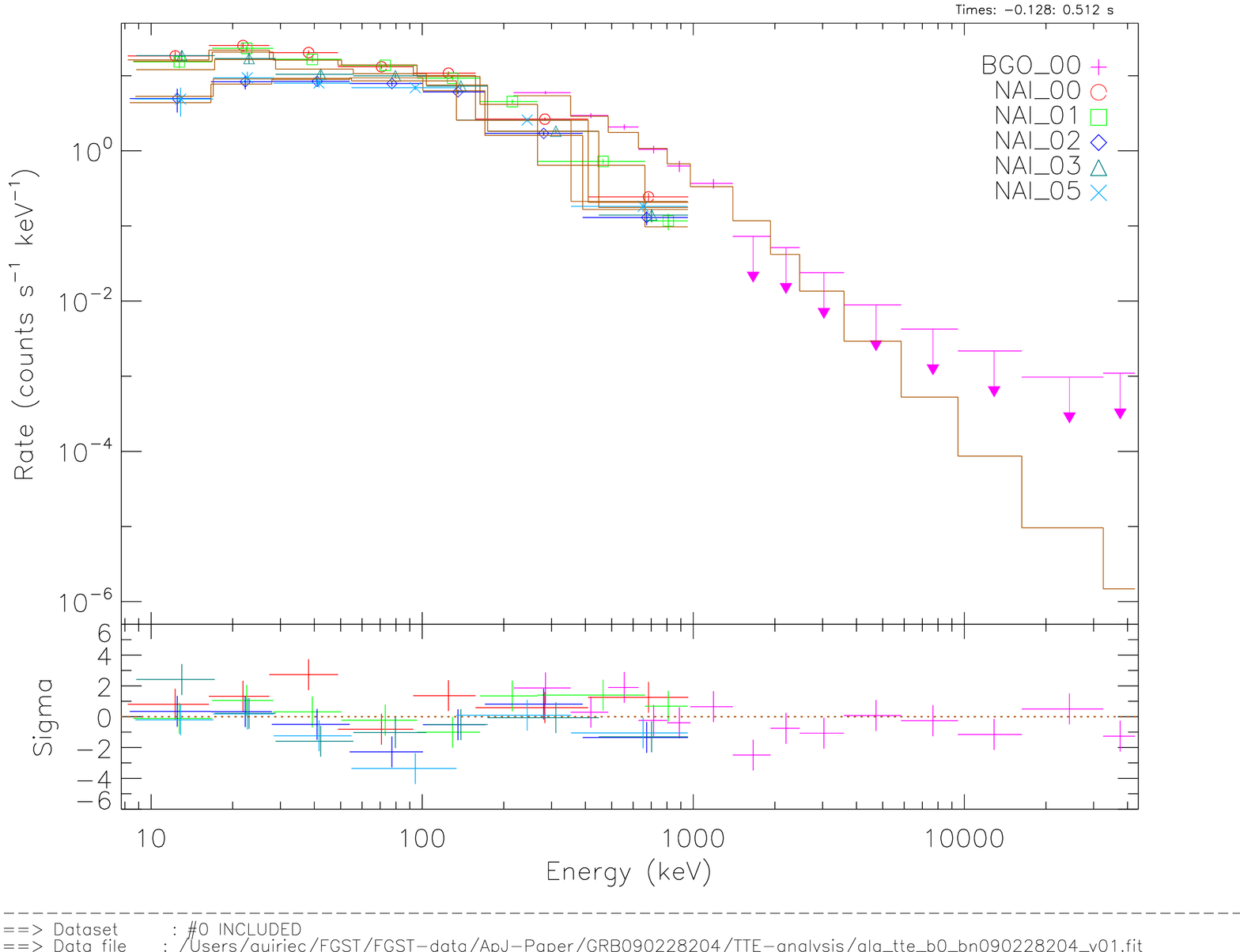}
\includegraphics[totalheight=0.17\textheight, viewport=8 0 540 730,clip,angle=90]{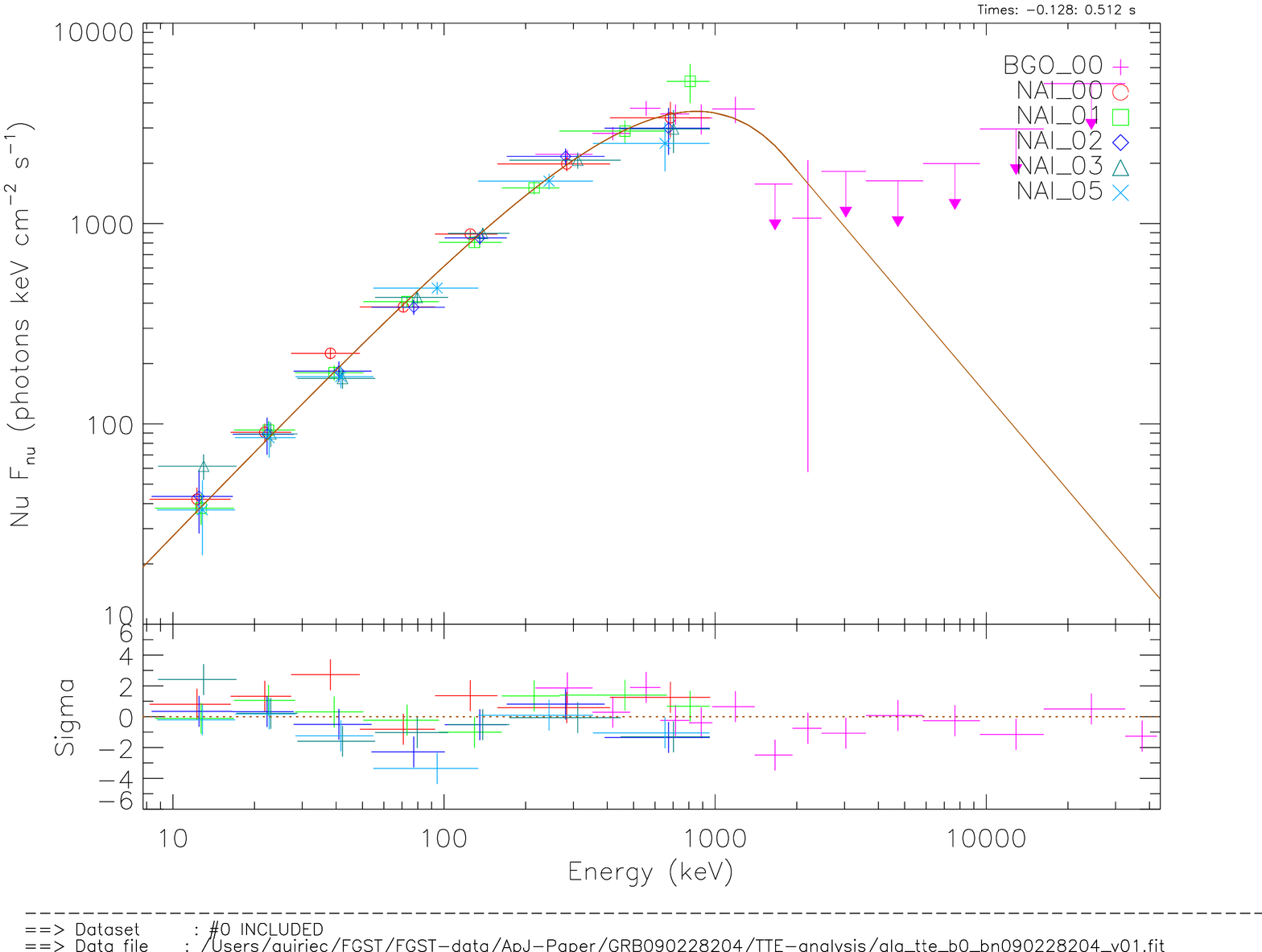}

\includegraphics[totalheight=0.17\textheight, viewport=8 0 540 730,clip,angle=90]{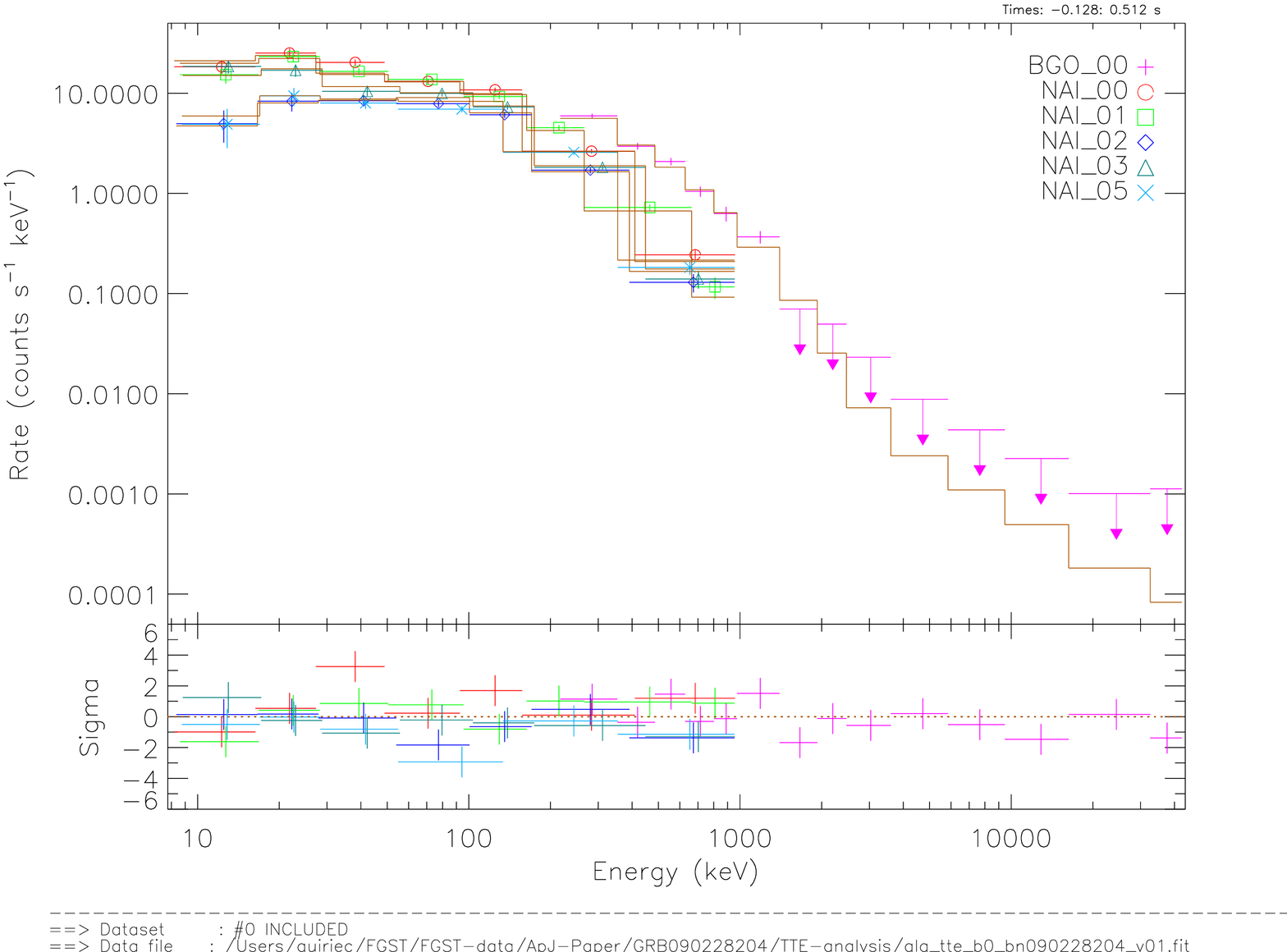}
\includegraphics[totalheight=0.17\textheight, viewport=8 0 540 730,clip,angle=90]{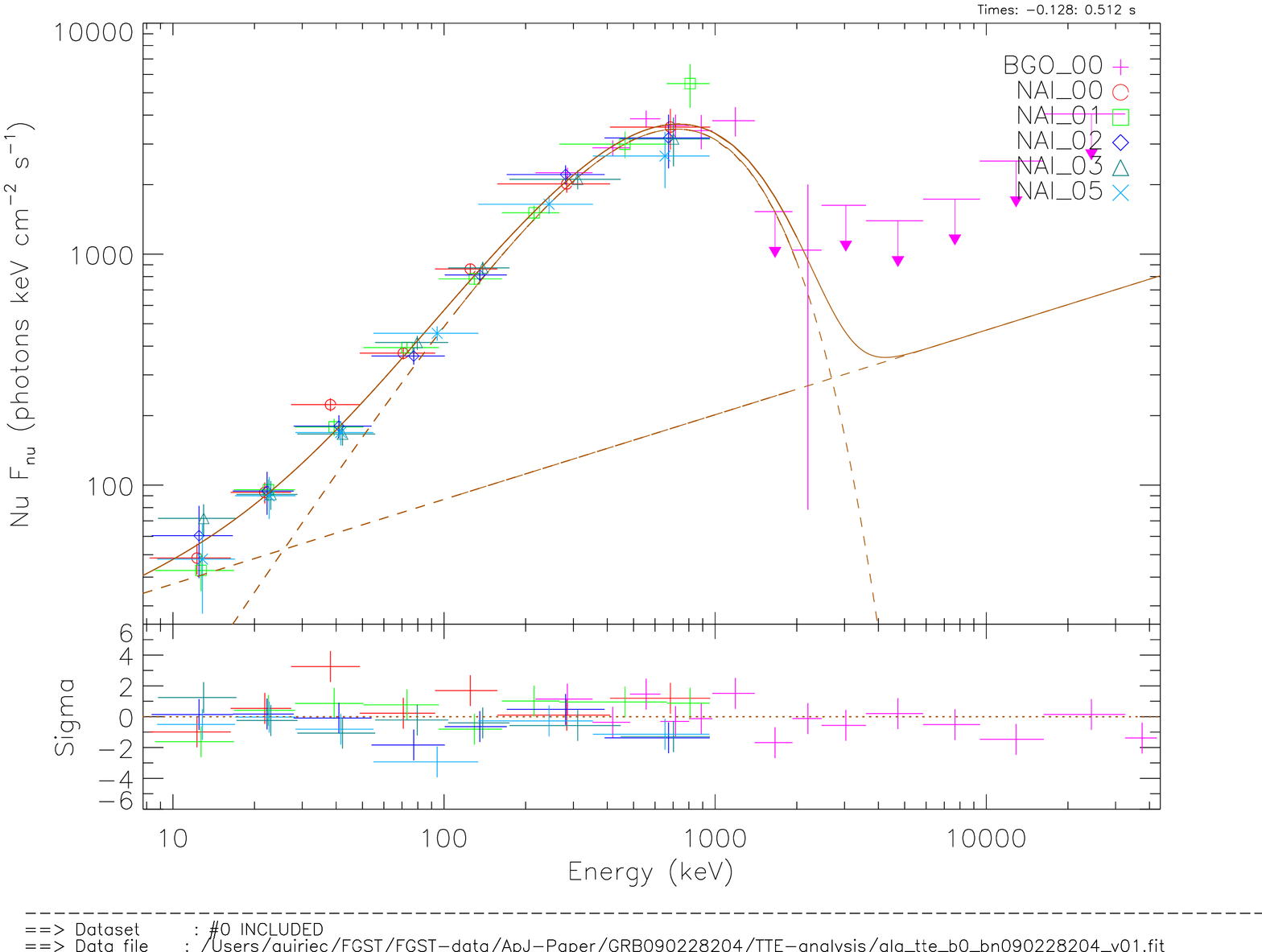}

\includegraphics[totalheight=0.17\textheight, viewport=8 0 540 730,clip,angle=90]{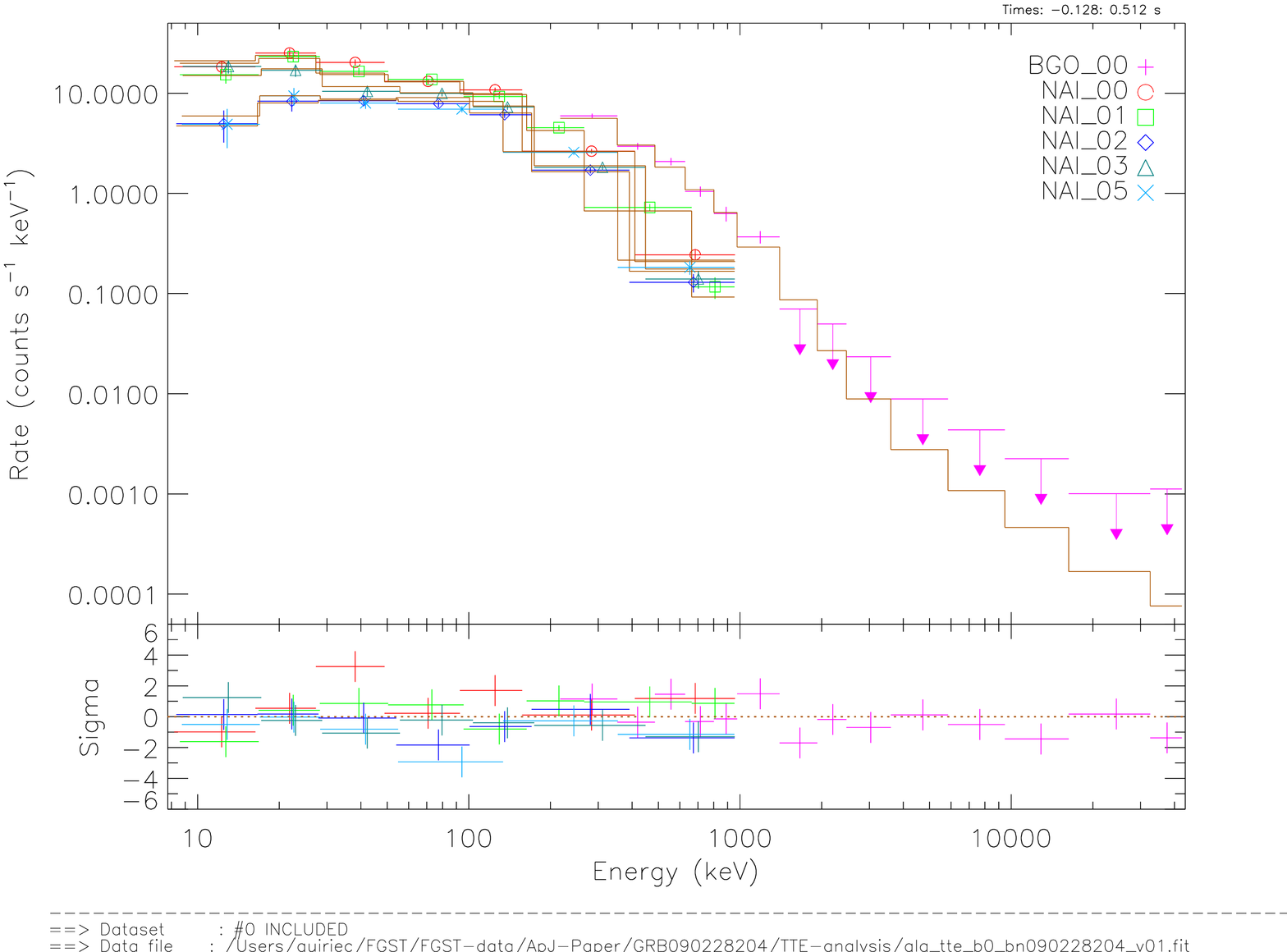}
\includegraphics[totalheight=0.17\textheight, viewport=8 0 540 730,clip,angle=90]{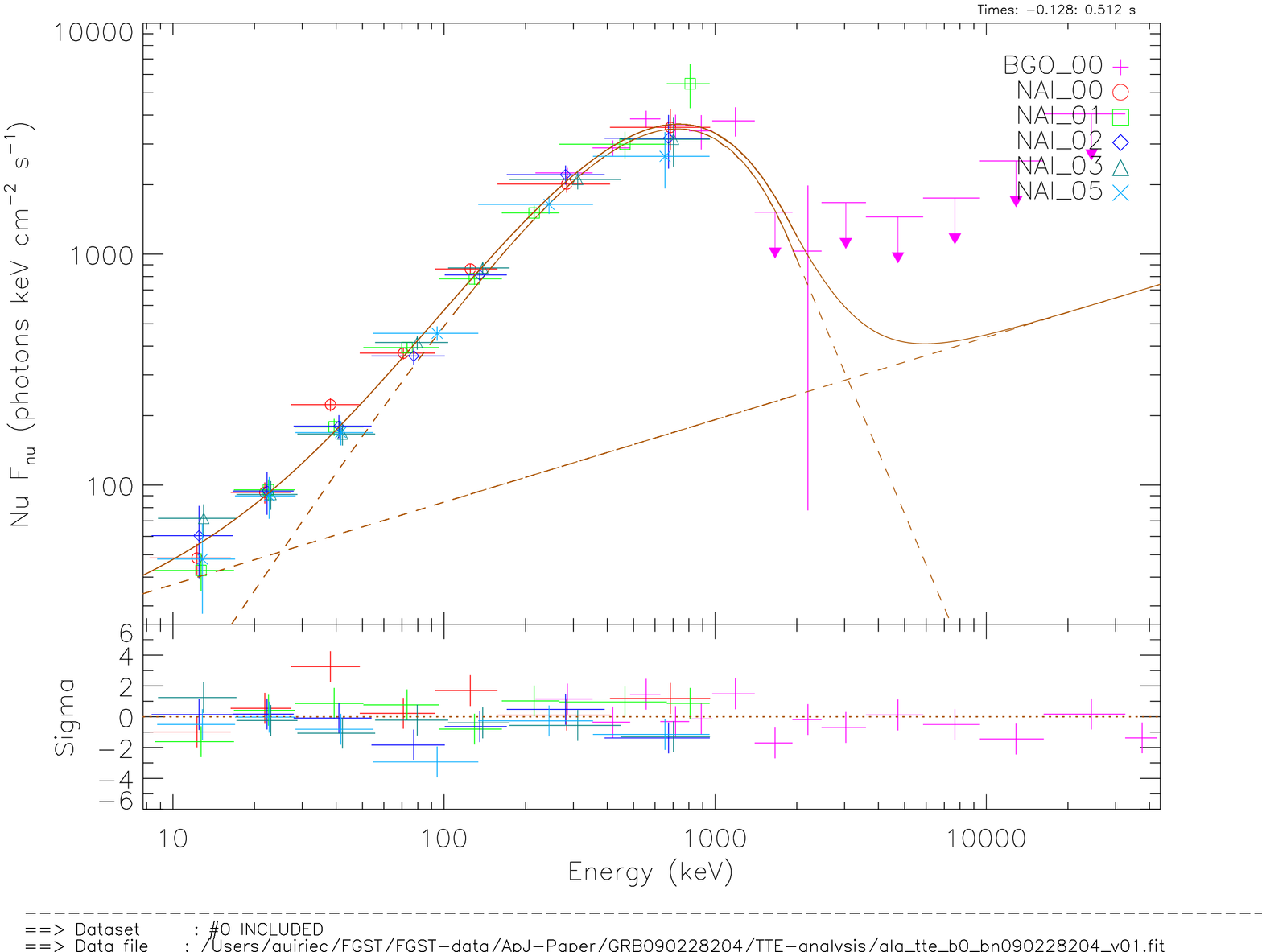}
\caption{Integrated count spectra ({\it left column}) and ${\nu}F_{\nu}$ spectra ({\it right column}) for GRB~$090228$ with Band ($1^{st}$ row), Compt+PL ($2^{nd}$ row) and Band+PL ($3^{rd}$ row) spectral fits.\label{fig:GRB090228_integrated_spectra}}
\end{figure}

\begin{figure}
\includegraphics[totalheight=0.17\textheight, viewport=8 0 540 730,clip,angle=90]{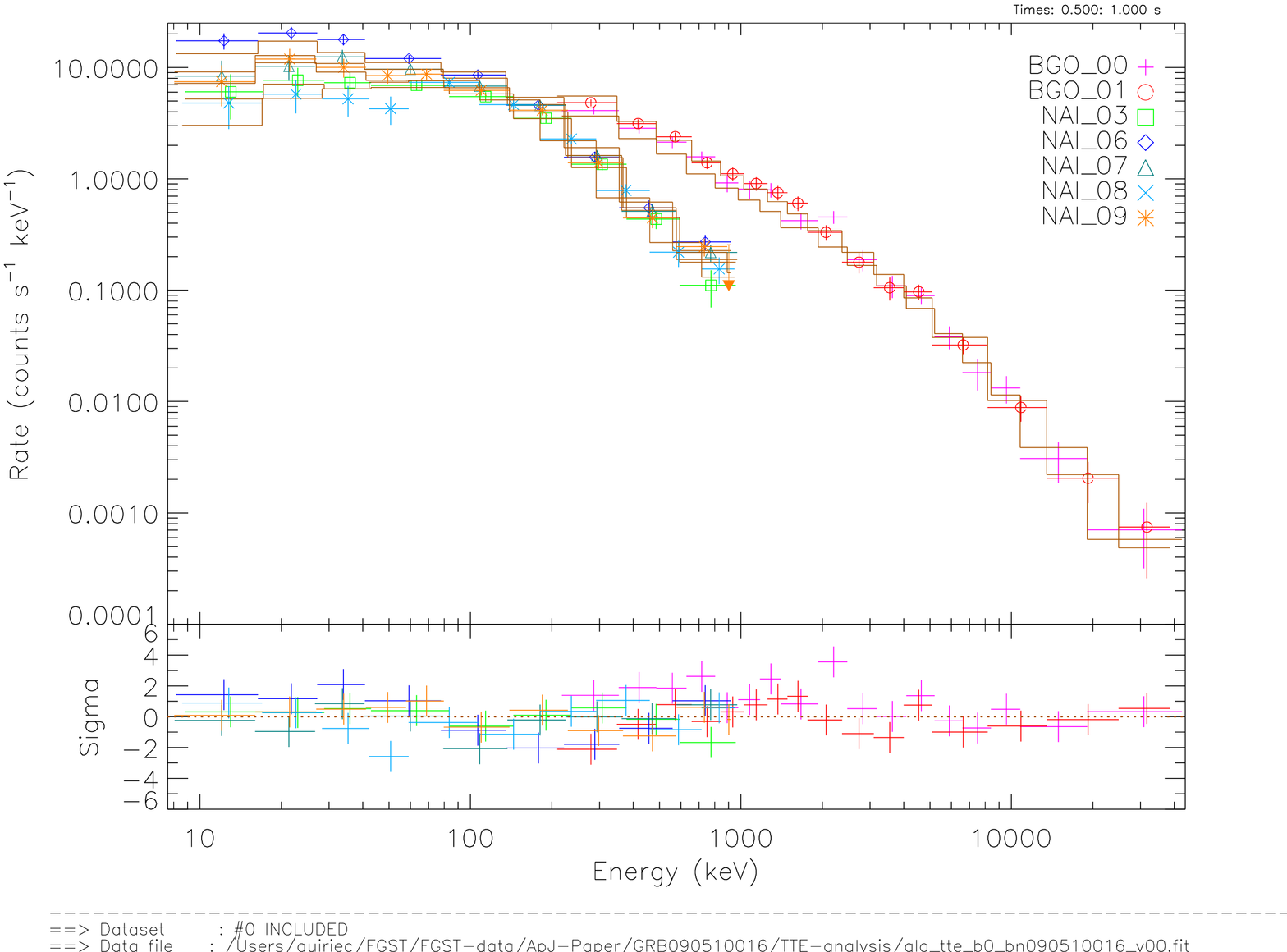}
\includegraphics[totalheight=0.17\textheight, viewport=8 0 540 730,clip,angle=90]{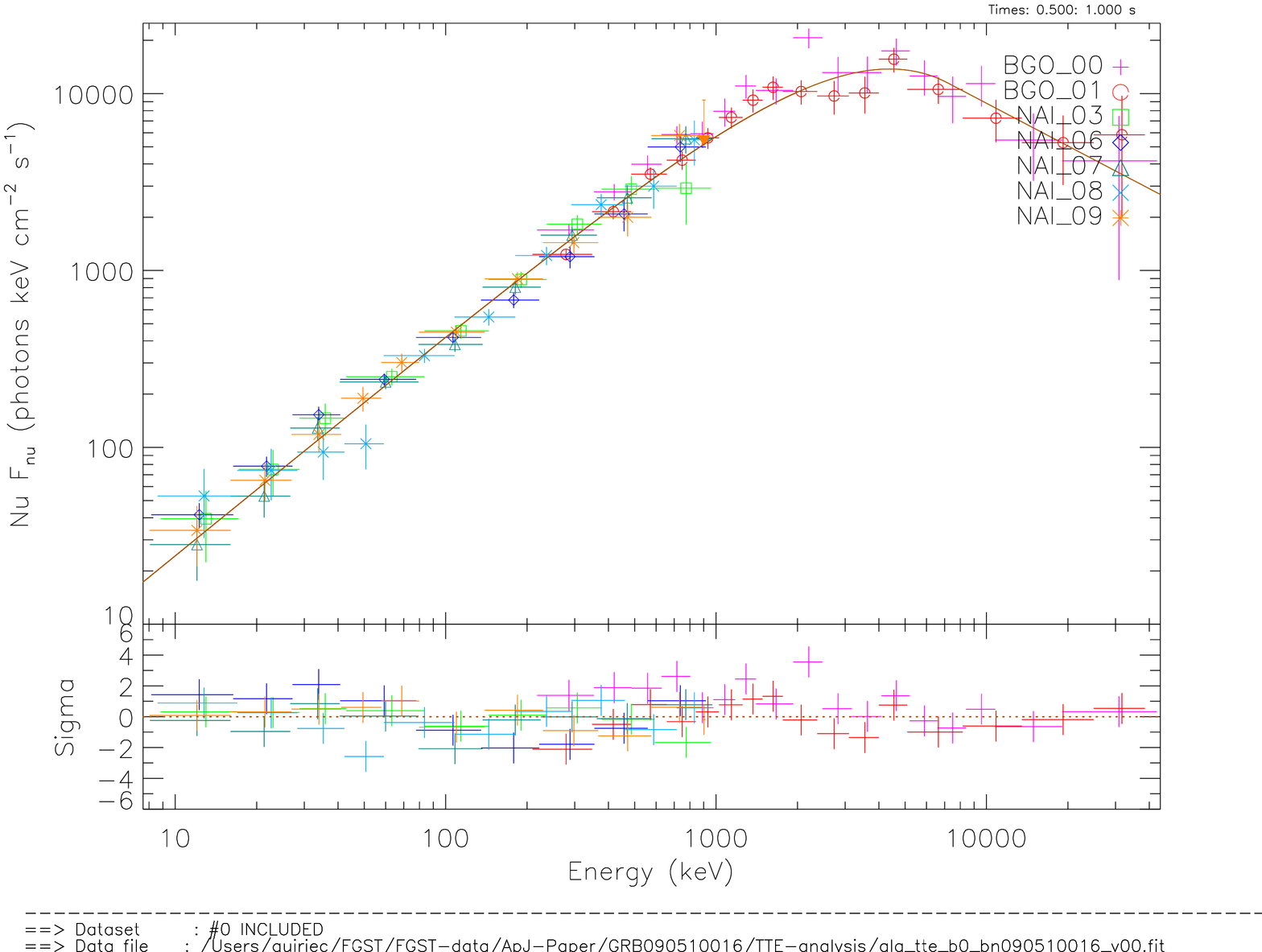}

\includegraphics[totalheight=0.17\textheight, viewport=8 0 540 730,clip,angle=90]{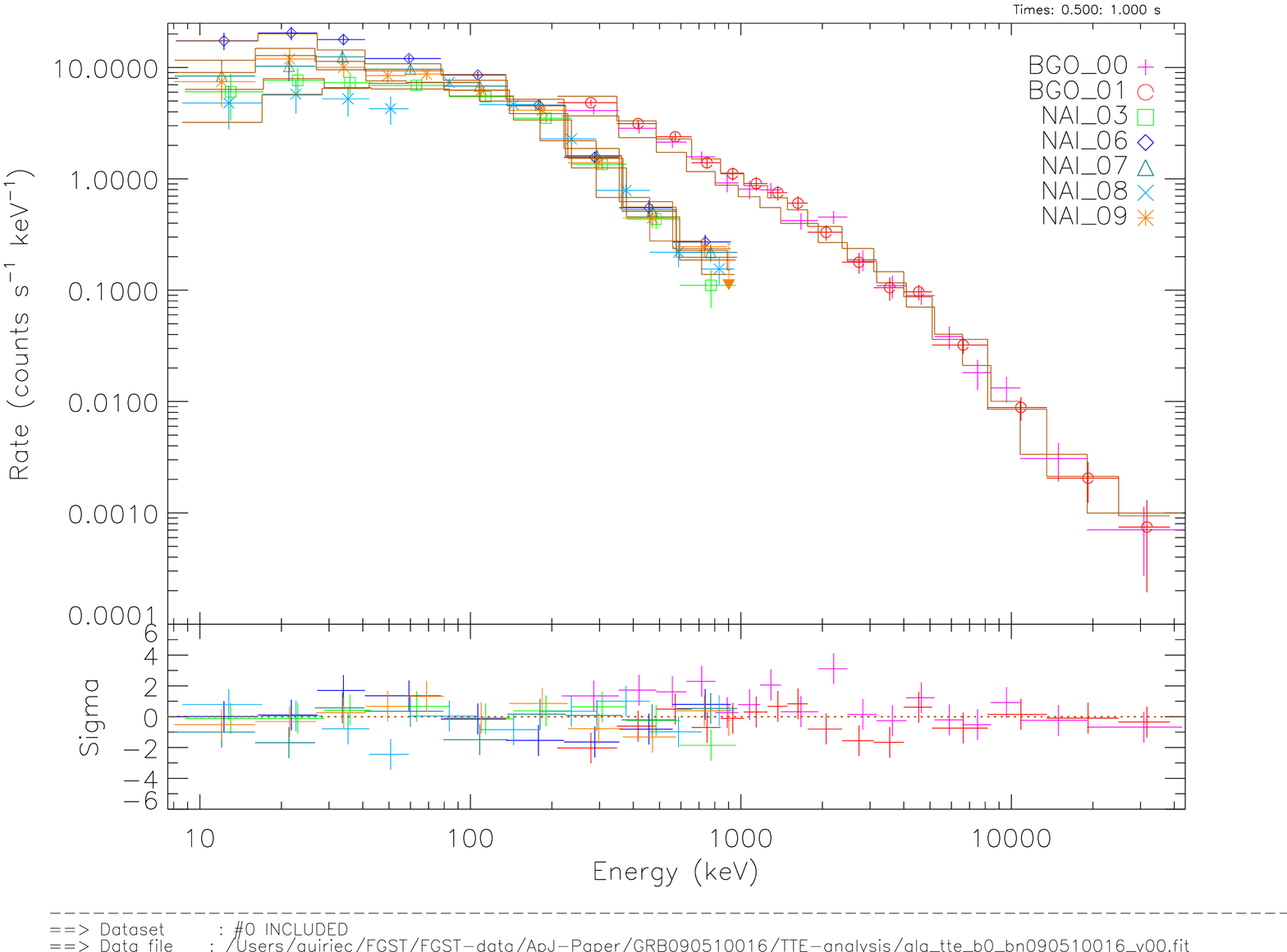}
\includegraphics[totalheight=0.17\textheight, viewport=8 0 540 730,clip,angle=90]{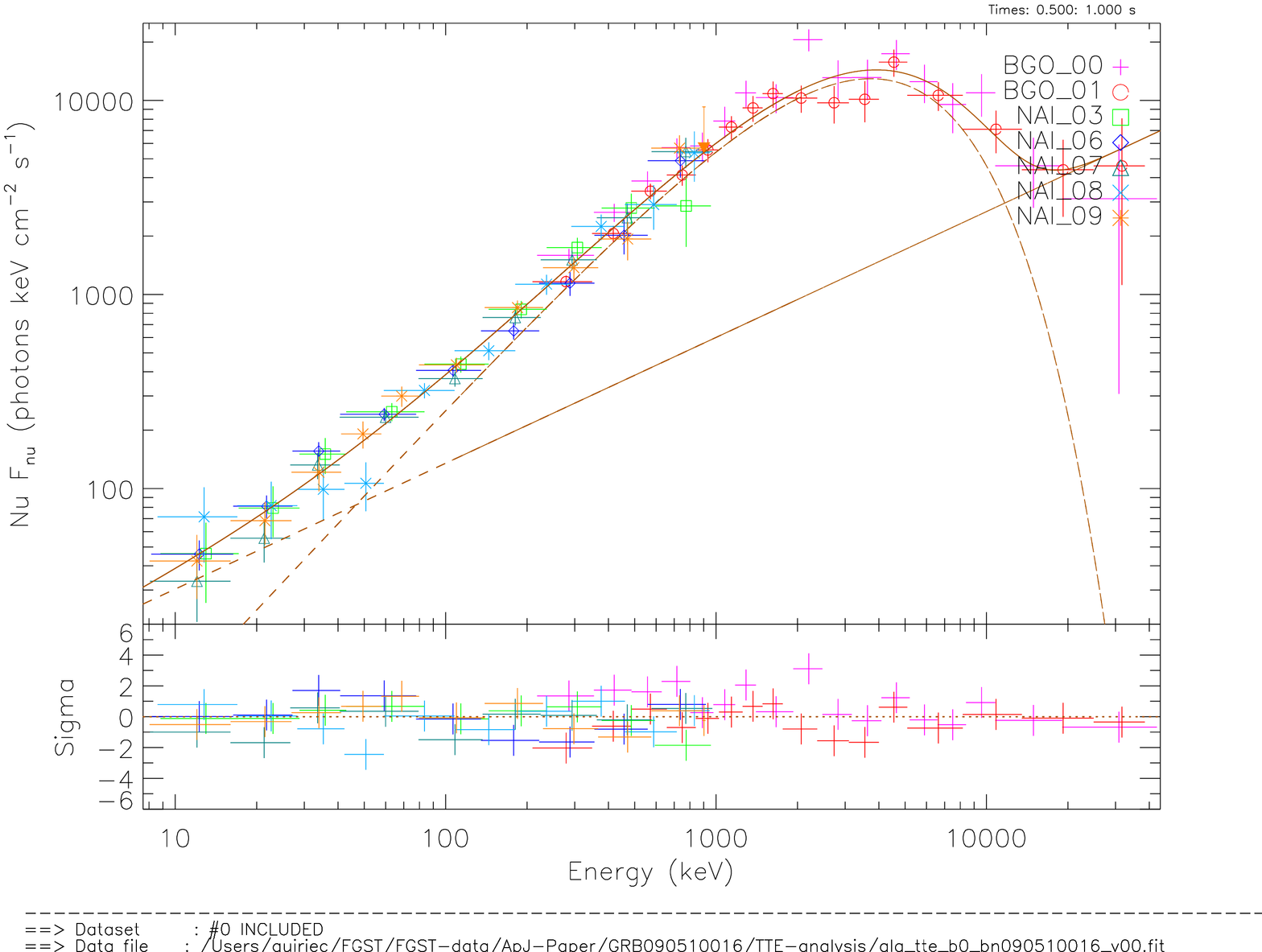}

\includegraphics[totalheight=0.17\textheight, viewport=8 0 540 730,clip,angle=90]{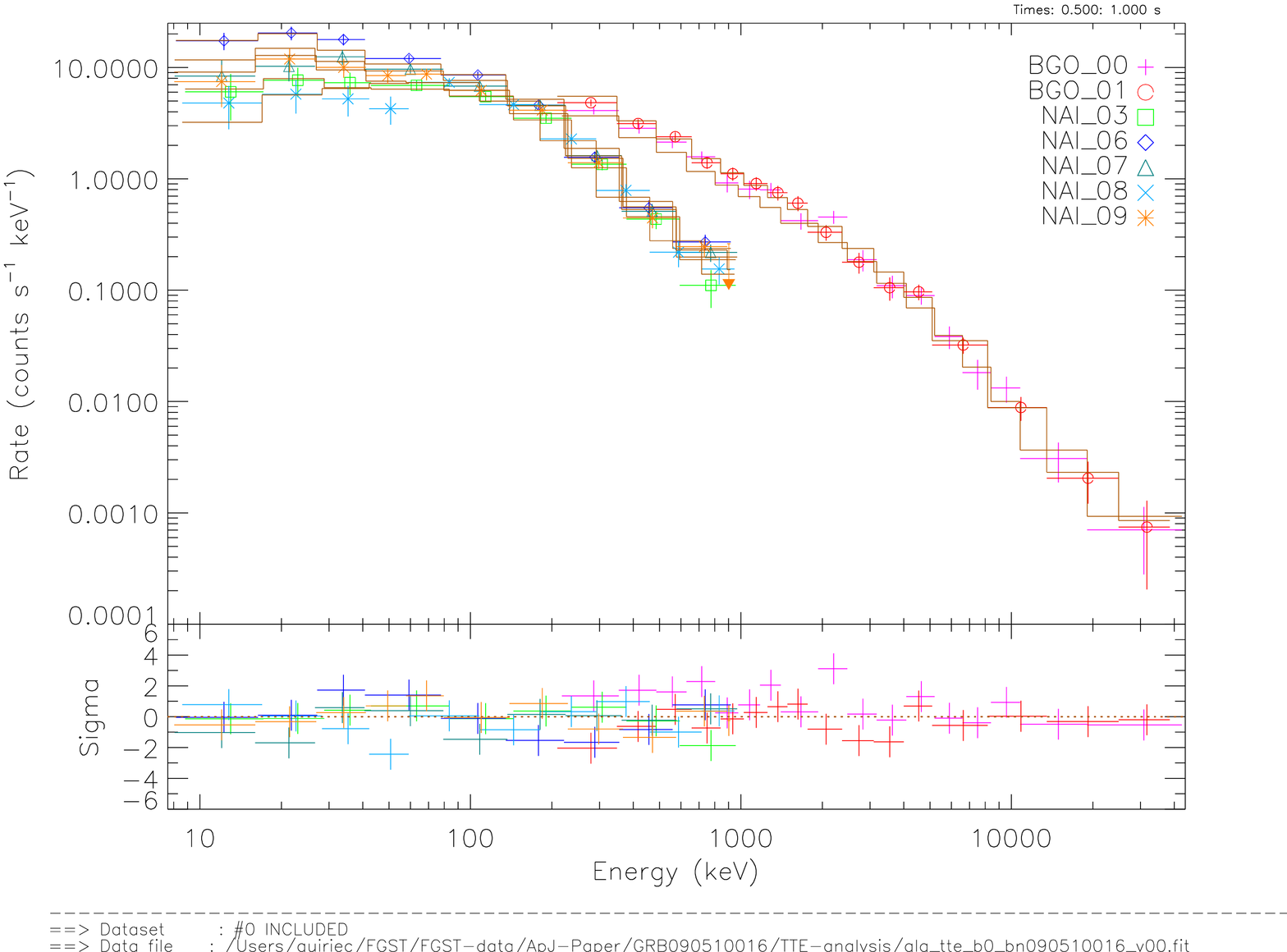}
\includegraphics[totalheight=0.17\textheight, viewport=8 0 540 730,clip,angle=90]{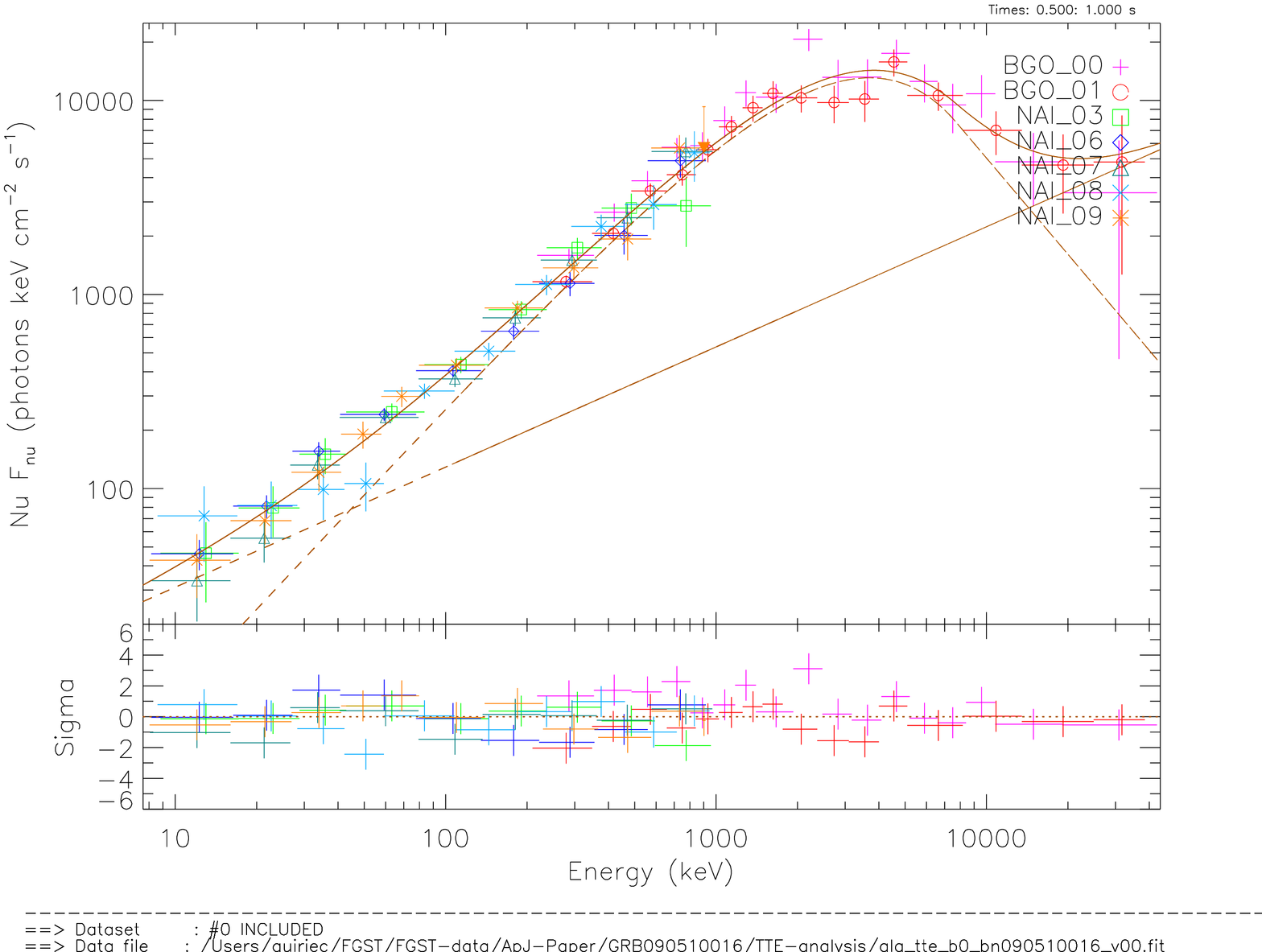}
\caption{Integrated count spectra ({\it left column}) and ${\nu}F_{\nu}$ spectra ({\it right column}) for GRB~$090510$ with Band ($1^{st}$ row), Compt+PL ($2^{nd}$ row) and Band+PL ($3^{rd}$ row) spectral fits.\label{fig:GRB090510_integrated_spectra}}
\end{figure}

Figures~\ref{fig:GRB090227B_integrated_spectra}~to~\ref{fig:GRB090510_integrated_spectra} show, for each of these GRBs, spectral fits for Band, Compt+PL and Band+PL, which all provided better fits than Compt (not shown). The power-law component appears to overpower the standard spectral component below few tens of keV 
and above few MeV, though most of the statistical significance is attributable to the fit of the low energy excess. The indices of the 
power laws have similar values ($\sim-$1.5) for all events, and they are also similar to the value found in the joint GBM+LAT fit to 
GRB~$090510$ already reported in~\citet{Abdo:2009,Ackermann:2010:GRB090510}~(Table~\ref{table:integrated_spectra}). While there are 
counts in GRBs~$090227$B and~$090510$ above the transition of the standard function to the additional power law at high energies, 
in GRB~$090228$, this additional component is only defined by counts below this transition.

Interestingly, the lag analysis presented in section~\ref{subsection:lightcurve} shows a significant delay of the high energy gamma rays above several MeV compared to the low energy gamma rays in the case of GRB~$090510$ and a weak trend in the case of GRB~$090227$B, in agreement with the spectral deviation from the Band function identified at high energies in these two GRBs. It may be that the positive lags identified at high energies in short GRBs are strong indicators for the existence of an additional spectral component. Similarly, the weak evidence of time lags measured between the 4--20 keV energy band and the other ones spreading from 20 keV to 8 MeV for GRBs~$090227$B and $090228$ could indicate the spectral deviation from the Band function observed at low energy.

An extra-component above several MeV has been reported for the first time in \cite{Gonzalez:2003} from observations of the long GRB~$941017$ with the Energetic Gamma Ray Experiment Telescope (EGRET) onboard {\it CGRO}. Since then, extra components have been observed several 
times in combined {\it FGST}/GBM+LAT observations of short and long GRBs, such as GRB~$090510$~\citep{Abdo:2009,Ackermann:2010:GRB090510} and GRB~$090902B$~\citep{Abdo:2009:GRB090902B}, where a spectral extension of this high-energy component at low energies has also been reported. The 
possible nature and implications of this extra component are discussed in section~\ref{section:interpretation}.

\begin{table*}

\caption{Fit parameters of the time-integrated spectra of GRB~$090227$B, GRB~$090228$ and GRB~$090510$ for 4 models: Band function (ie. Band), exponential cut-off power law (ie. Compt for Comptonized), Band+Power Law (ie Band+PL) and Compt+Power Law (ie. Compt+PL). The time-interval over which the fits are performed respectively $-$0.128 s to 0.384 s, $-$0.128 s to 0.512 s and 0.500 s to 1.000 s since the trigger time for GRB~$090227$B, GRB~$090228$ and GRB~$090510$. All the asymetrical errors are given at the 1-$\sigma$ confident level. The normalization values are computed at 100 keV for Compt and at 10 MeV for PL.\label{table:integrated_spectra}}

\begin{center}
\begin{tabular}{|l|l|lll|l|c|}
\hline
\multicolumn{1}{|c|}{Name} & \multicolumn{1}{|c|}{Model} & \multicolumn{3}{|c|}{Band or Compt} & \multicolumn{1}{|c|}{PL} & \multicolumn{1}{|c|}{Castor C-stat/d.o.f.}\\
\hline
\multicolumn{1}{|c|}{}&\multicolumn{1}{|c|}{} & \multicolumn{1}{|c}{$E_{peak}$ (keV)} & \multicolumn{1}{c}{$\alpha$ or index} & \multicolumn{1}{c|}{$\beta$} & \multicolumn{1}{|c|}{index} & \multicolumn{1}{|c|}{} \\
\hline
GRB~$090227$B & Band  & $2116\pm{97}$   & $-0.50\pm{0.02}$ & $-3.35_{-0.39}^{+0.27}$ & \multicolumn{1}{|c|}{n/a} & 699/607 \\
            & Compt   & $2227\pm{90}$   & $-0.52\pm{0.02}$ & \multicolumn{1}{c|}{n/a} & \multicolumn{1}{|c|}{n/a} & 706/608 \\
            & Band+PL & $1947_{-98}^{+205}$  & $-0.36_{-0.13}^{+0.05}$ & $-3.44_{-0.80}^{+0.58}$ & $-1.51\pm{0.05}$ & 686/605 \\
            & Compt+PL& $1995\pm{96}$ & $-0.36\pm{0.05}$ & \multicolumn{1}{c|}{n/a} & $-1.37\pm{0.06}$ & 689/606 \\
\hline
GRB~$090228$ & Band  & $860\pm{50}$   & $-0.59\pm{0.03}$ & $-3.77_{-6.64}^{+0.64}$& \multicolumn{1}{|c|}{n/a} & 813/728 \\
           & Compt   & $862\pm{52}$   & $-0.59\pm{0.03}$ & \multicolumn{1}{c|}{n/a} & \multicolumn{1}{|c|}{n/a} & 813/729 \\
           & Band+PL & $723\pm{45}$   & $-0.24\pm{0.10}$ & $-4.74_{-\infty}^{+1.14}$& $-1.64\pm{0.03}$ & 795/726 \\
           & Compt+PL& $722\pm{47}$   & $-0.23\pm{0.11}$ & \multicolumn{1}{c|}{n/a} & $-1.63_{-0.15}^{+0.09}$ & 795/727 \\
\hline
GRB~$090510$ & Band  & $4383\pm{290}$ & $-0.75\pm{0.02}$ & $-2.80_{-0.28}^{+0.20}$ & \multicolumn{1}{|c|}{n/a} & 911/850 \\
           & Compt   & $4797_{-237}^{+255}$ & $-0.77\pm{0.02}$ & \multicolumn{1}{c|}{n/a} & \multicolumn{1}{|c|}{n/a} & 922/851 \\
           & Band+PL & $3695\pm{248}$ & $-0.51\pm{0.08}$ & $-3.65_{-\infty}^{+0.75}$ & $-1.38\pm{0.04}$ & 897/848 \\
           & Compt+PL& $3731\pm{265}$ & $-0.58\pm{0.07}$   & \multicolumn{1}{c|}{n/a} & $-1.35\pm{0.04}$  & 897/849 \\
 (GBM+LAT) & Band+PL & $3936\pm{280}$ & $-0.58\pm{0.06}$ & $-2.83\pm{0.20}$ & $-1.62\pm{0.03}$ & \\
\hline
\end{tabular}
\end{center}
\end{table*}

\subsubsection{Comparison of the spectral results with the literature}
\label{section:standard_parameters}

It has long been known that short GRBs are harder than long GRBs, but there has been a debate about the manner in which they are harder.~\citet{Paciesas:2003}, using a {\it BATSE} datatype with coarse temporal (2 s) and spectral (16 channels) resolution, found that short GRBs were harder in all spectral parameters. When a Compt function was used, both the index and $E_{\rm peak}$ were harder than the ones for the long GRBs; when the Band function was used, both indices and $E_{\rm peak}$ were harder. In contrast,~\citet{Ghirlanda:2004,Ghirlanda:2009b} found that the higher hardness of short GRBs is entirely due to their harder low-energy indices, while their $E_{\rm peak}$ distributions are similar to long GRBs. 

The $E_{\rm peak}$ values of our best time-integrated fits (Compt+PL) in all events are very high: $2000 \pm100$ keV (GRB~$090227$B), $720 \pm50$ keV (GRB~$090228$) and $3730 \pm270$ keV (GRB~$090510$). The highest two values are above any of the other 320 GRBs detected with GBM from July 2008 to October 2009, and in fact higher than any $E_{\rm peak}$ value reported thus far with the possible exception of the {\it Konus/Wind} online catalog of short 
GRBs~\citep[][, see also http://www.ioffe.ru/LEA/shortGRBs/Catalog/]{Mazets:2004}. About 25\% of the prompt emission spectra for 86 short GRBs with acceptably constrained fit parameters in the {\it Konus} catalog have $E_{\rm peak}$ values, which could be above 1 MeV within the error bars, but all are below 2.5 MeV. Our GBM results show that some bright short GRBs have exceptional $E_{\rm peak}$ values, beyond the range observed in long GRBs. That  Epeak values for long GRBs end at $\sim$1 MeV is supported by the analysis of Solar Maximum Mission ({\it SMM}) data by~\citet{Harris:1998}, who performed a dedicated GRB search to address this question. They searched for hard-spectrum long GRBs in the 0.8--10 MeV range and only 2 were found in the 0.8 - 10 MeV range which were not part of the 177 burst sample detected in the 0.35 - 0.8 MeV range. They concluded that there is no population of hard-spectrum long GRBs hidden by instrumental effects by {\it SMM}, {\it BATSE}, and other instruments which triggered on low-energy gamma-rays. The analysis of~\citet{Paciesas:2003} did not reveal such extreme values in short GRBs because of the limited {\it BATSE}/LAD spectral resolution and energy coverage ($<2$ MeV). An independent analysis of the same {\it BATSE} data~\citep{Ghirlanda:2004,Ghirlanda:2009b} found a narrow distribution of low $E_{\rm peak}$ values for all GRBs with fluxes below $4\times 10^{-6}$ erg.cm$^{-2}$.s$^{-1}$, and that short GRBs are spectrally similar to the first one or two seconds of long GRBs. Our (limited) sample of three short GRBs with fluences above $2 \times 10^{-6}$ erg.cm$^{-2}$ does not support their results. 

Previous analyses of short GRBs have typically used a cutoff power law for most fits in preference to the Band function commonly used for long GRBs~\citep[][ http://www.ioffe.ru/LEA/shortGRBs/Catalog/]{Ghirlanda:2004,Ghirlanda:2009b,Mazets:2004}. It has not been clear whether the inability to show that a Band function is preferred over the cutoff power law indicates an intrinsic difference between short and long GRBs, or merely a lack of sufficient counts in short GRBs~\citep{Ghirlanda:2009}. This issue is best addressed with the broad energy coverage of the GBM detectors and our selection of a sample that is extremely bright, which should enable better determination of the high energy power-law index, $\beta$, of the Band function. However, we have, as in previous experiments, found it difficult to constrain $\beta$. We suggest that this difficulty has 3 main causes: (i) $E_{\rm peak}$ is higher in short bursts than in long bursts, leaving the measurement of $\beta$ to be made at higher 
energies where there are fewer counts, (ii) $\beta$ is intrinsically very steep, often resulting in a lack of preference for a power-law 
over an exponential cut-off to the spectrum above $E_{\rm peak}$, and (iii) the presence of an underlying power-law component in addition to the 
standard Band function makes $\beta$ difficult to quantify. We conclude that we cannot resolve this question using a single spectral component, but instead that an additional power-law component is statistically preferred. The Band+PL model is not statistically required, compared to Compt+PL, but using this model allows us to measure $\beta$ or at least to measure upper limits (Table~\ref{table:integrated_spectra}). These values confirm the steepness of $\beta$: $-3.44^{+0.58}_{-0.8}$ for GRB~$090227$B, $<-3.60$ for GRB~$090228$ and $<-2.90$ for GRB~$090510$.

Finally, the GBM results for the time-integrated low-energy index, $\alpha$, are consistent with previous measurements~\citep{Paciesas:2003,Ghirlanda:2004,Ghirlanda:2009b}, and similar to about half of the~\citet{Mazets:2004} catalog of short bursts, with $\alpha$ harder than $-$0.67. We conclude here that both $\alpha$ and $E_{\rm peak}$ contribute to the hardness of short GRBs, but the steep high-energy spectral-indices indicate a soft spectrum above the break energy. 

\subsection{Time-Resolved Spectral Analysis}

Dividing the data into fine time intervals reduces the number of counts in each spectrum and it is not possible to fit models as complex as were used for the time-integrated analysis. We selected our time bins to be as fine as possible, while still allowing statistically significant spectral analysis by having enough counts distributed over the energy channels; the resolution ranges from 2~ms to 94~ms. We fit our time-resolved spectra only with the Band function since the statistical content in small time bins is too low to measure the effect of the extra spectral component. Using Band only instead of Band+PL artificially makes the high-energy index of the Band function softer and $E_{\rm peak}$ slightly harder. However, the effects on the spectral parameters are negligible, and in the time bins where Band+PL can be fit, we find that the systematic discrepancies between the Band and Band+PL fit values are smaller than the statistical errors. In many cases, we find $\beta$ consistent with $-\infty$ and provide only $1\sigma$ upper-limits. The fits parameters for each time interval are presented in Tables~\ref{table:time_resolved_spectroscopy_GRB090227B},~\ref{table:time_resolved_spectroscopy_GRB090228}~and~\ref{table:time_resolved_spectroscopy_GRB090510}
for GRBs~$090227$B,~$090228$~and~$090510$, respectively.

In the past, we have automatically determined time intervals by accumulating data until a signal-to-noise ratio requirement is reached, but this method merges peaks and valleys of the light curves, which we strongly wish to avoid. In addition, the number of counts required to perform a good fit depends on the spectral shape. Spectra spread over a large energy range (such as for hard spectra) require more counts to have a good fit than spectra covering a smaller energy range (soft spectra). We have selected statistically significant and variable time bins for our time-resolved spectra to account for the various pulse structures in the 2 ms light curves. This results in time bins containing either peaks or valleys of the light curves but not both when possible. To evaluate the reliability of the spectral fits in these time bins, we performed simulations for several low fluence spectra with low and high $E_{\rm peak}$ values as well as with steep $\beta$. The simulated sets included backgrounds from the real data; the signal was generated using the model function of the  real data fit. This input function was folded through the GBM NaI and BGO responses giving a source count spectrum as would be observed with the detectors. Poissonian fluctuations were then applied to each energy channel of the sum of the source and background for each simulated spectrum. Each spectral simulation set included ten thousand spectra which were then fit with the input function model using RMFIT. We found that all fit parameters were correctly reconstructed with good constraints and with a systematic bias smaller than the statistical errors. 

\begin{table}[h!]
\caption{Parameters of the Band function for the fine-time resolved spectroscopy of GRB~$090227$B. All the assymetrical errors are given at the 1-$\sigma$ confident level. When the high-energy spectral index $\beta$ cannot be constrained, the 1-$\sigma$ upper limit is reported.\label{table:time_resolved_spectroscopy_GRB090227B}}
\begin{center}
{\tiny
\begin{tabular}{|ll|lll|}
\hline
\multicolumn{1}{|c}{Tstart} & \multicolumn{1}{c|}{Tstop} & \multicolumn{1}{c}{$E_{peak}$~(keV)}  & \multicolumn{1}{c}{$\alpha$} & \multicolumn{1}{c|}{$\beta$}\\
\hline
-0.022 & -0.012 & $298_{-57}^{+81}$    & $0.09_{-0.38}^{+0.55}$ & $<-2.85$                \\
-0.012 & -0.002 & $1690_{-169}^{+211}$ & $-0.29\pm{0.09}$       & $<-4.30$                \\
-0.002 &  0.004 & $1140_{-140}^{+157}$   & $0.18_{-0.16}^{+0.20}$ & $<-3.16$                \\
 0.004 &  0.010 & $1891_{-169}^{+197}$ & $0.02\pm{0.11}$        & $<-4.10$                \\
 0.010 &  0.012 & $784_{-166}^{+208}$  & $1.83_{-0.81}^{+1.82}$  & $-2.41_{-0.49}^{+0.30}$ \\
 0.012 &  0.016 & $2347_{-242}^{+292}$ & $0.11\pm{0.14}$        & $<-3.48$                \\
 0.016 &  0.018 & $1314_{-288}^{+231}$ & $0.71_{-0.32}^{+0.67}$ & $<-2.88$                \\
 0.018 &  0.022 & $2139_{-369}^{+527}$ & $0.01\pm{0.16}$        & $<-2.38$                \\
 0.022 &  0.026 & $1770_{-173}^{+206}$ & $0.15\pm{0.15}$        & $<-4.05$                \\
 0.026 &  0.028 & $1387_{-187}^{+244}$ & $0.29\pm{0.29}$        & $<-3.91$                \\
 0.028 &  0.032 & $2129_{-312}^{+437}$ & $0.03\pm{0.18}$        & $<-2.99$                \\
 0.032 &  0.034 & $2741_{-544}^{+599}$ & $0.18_{-0.22}^{+0.33}$ & $<-2.69$                \\
 0.034 &  0.038 & $2316_{-286}^{+351}$ & $-0.18\pm{0.12}$       & $<-3.45$                \\
 0.038 &  0.040 & $3222_{-696}^{+1030}$& $-0.71\pm{0.10}$       & $<-2.75$                \\
 0.040 &  0.044 & $2482_{-370}^{+436}$ & $-0.32\pm{0.13}$       & $<-3.55$                \\
 0.044 &  0.048 & $2040_{-373}^{+452}$ & $-0.15\pm{0.19}$       & $-3.50_{-2.79}^{+0.83}$ \\
 0.048 &  0.050 & $2916_{-903}^{+1080}$& $-0.51\pm{0.19}$       & $-2.79_{-1.53}^{+0.71}$ \\
 0.050 &  0.058 & $1711_{-307}^{+404}$ & $-0.18\pm{0.16}$       & $-2.32_{-0.35}^{+0.25}$ \\
 0.058 &  0.076 & $1399_{-232}^{+291}$ & $-0.24\pm{0.12}$       & $-2.49_{-0.59}^{+0.28}$ \\
 0.076 &  0.090 & $1710_{-197}^{+276}$ & $-0.47\pm{0.08}$       & $<-2.92$                \\
 0.090 &  0.104 & $741_{-143}^{+176}$  & $-0.17\pm{0.17}$       & $-2.37_{-0.59}^{+0.29}$ \\
 0.104 &  0.112 & $2301_{-327}^{+417}$ & $-0.53\pm{0.08}$       & $<-3.4$                 \\
 0.112 &  0.118 & $408_{-126}^{+174}$  & $0.02_{-0.31}^{+0.45}$ & $1.96_{-0.31}^{+0.21}$  \\
 0.118 &  0.120 & $1639_{-505}^{+810}$ & $-0.52\pm{0.20}$       & $<-2.15$                \\
 0.120 &  0.126 & $403_{-101}^{146}$   & $-0.07\pm{0.04}$       & $<1.74$                 \\
 0.126 &  0.142 & $189_{-32}^{+40}$    & $0.12_{-0.23}^{+0.65}$ & $<-2.36$                \\
 0.142 &  0.188 & $101_{-27}^{+34}$    & $-0.40_{-0.37}^{+0.58}$& $-2.00_{-0.56}^{+0.22}$ \\
\hline
\end{tabular}
}
\end{center}
\end{table}

\begin{table}[h!]
\caption{Parameters of the Band function for the fine-time resolved spectroscopy of GRB~$090228$. All the assymetrical errors are given at the 1-$\sigma$ confident level. When the high-energy spectral index $\beta$ cannot be constrained, the 1-$\sigma$ upper limit is reported.\label{table:time_resolved_spectroscopy_GRB090228}}
\begin{center}
{\tiny
\begin{tabular}{|ll|lll|}
\hline
\multicolumn{1}{|c}{Tstart} & \multicolumn{1}{c|}{Tstop} & \multicolumn{1}{c}{$E_{peak}$~(keV)}  & \multicolumn{1}{c}{$\alpha$} & \multicolumn{1}{c|}{$\beta$}\\
\hline
-0.008    &  0.012   &  $153\pm{41}$    &  $+0.29_{-0.43}^{+0.68}$  &  $-2.39_{-0.60}^{+0.33}$  \\
0.012     &  0.020   &  $1353\pm{233}$  & $-0.61\pm{0.08}$   & $<-3.28$  \\
0.020     &  0.022   &  $1497_{-232}^{+295}$  & $+0.07\pm{0.22}$   & $<-3.60$  \\
0.022     &  0.024   &  $1087\pm{237}$  & $+0.02\pm{0.22}$   & $-2.46_{-0.85}^{+0.35}$  \\
0.024     &  0.026   &  $857_{-150}^{+198}$   & $+0.09\pm{0.26}$   & $-2.89_{-0.88}^{+0.47}$  \\
0.026     &  0.028   &  $653_{-118}^{+195}$   & $+0.31_{-0.33}^{+0.42}$   & $-2.46_{-0.62}^{+0.32}$  \\
0.028     &  0.032   &  $738\pm{111}$   & $+0.08\pm{0.19}$   & $-3.56_{-2.90}^{+0.78}$  \\
0.032     &  0.036   &  $1046\pm{155}$  & $-0.01\pm{0.16}$   & $<-3.97$  \\
0.036     &  0.042   &  $711\pm{168}$   & $-0.09_{-0.18}^{+0.27}$   & $-2.94\pm{2.68}$  \\
0.042     &  0.048   &  $623_{-89}^{+107}$    & $-0.21\pm{0.17}$   & $<-2.65$  \\
0.048     &  0.054   &  $575\pm{88}$     & $+0.05\pm{0.24}$   & $<-3.08$  \\
0.054     &  0.060   &  $479_{-59}^{+75}$     & $-0.03\pm{0.22}$   & $<-3.45$  \\
0.060     &  0.066   &  $597\pm{76}$     & $+0.31\pm{0.23}$   & $<-3.93$  \\
0.066     &  0.072   &  $728_{-91}^{+115}$    & $-0.03\pm{0.18}$   & $<-2.99$  \\
0.072     &  0.078   &  $700\pm{119}$   & $+0.02\pm{0.20}$   & $<-2.56$  \\
0.078     &  0.082   &  $451_{-77}^{+107}$    & $+0.63\pm{0.48}$   & $-3.02_{-2.73}^{+0.61}$  \\
0.082     &  0.092   &  $650\pm{94}$     & $+0.09\pm{0.20}$   & $<-3.01$  \\
0.092     &  0.128   &  $307\pm{35}$     & $-0.14\pm{0.21}$   & $<-3.19$  \\
0.128     &  0.170   &  $42\pm{14}$       & $+0.92_{-1.67}^{+2.41}$   & $-2.92_{-0.83}^{+0.45}$  \\
0.170     &  0.212   &  $23\pm{24}$      & $-1.55_{-0.46}^{+0.00}$   & $<-2.68$  \\
\hline
\end{tabular}
}
\end{center}
\end{table}

\begin{table}[h!]
\caption{Parameters of the Band function for the fine-time resolved spectroscopy of GRB~$090510$. All the assymetrical errors are given at the 1-$\sigma$ confident level. When the high-energy spectral index $\beta$ cannot be constrained, the 1-$\sigma$ upper limit is reported.\label{table:time_resolved_spectroscopy_GRB090510}}
\begin{center}
{\tiny
\begin{tabular}{|ll|lll|}
\hline
\multicolumn{1}{|c}{Tstart} & \multicolumn{1}{c|}{Tstop} & \multicolumn{1}{c}{$E_{peak}$~(keV)}  & \multicolumn{1}{c}{$\alpha$} & \multicolumn{1}{c|}{$\beta$}\\
\hline
0.5      &  0.52    & $389_{-185}^{+530} $   &   $-0.68_{-0.43}^{+0.77}$     &  $<-2.26$    \\
0.52     &  0.53    & $390_{-123}^{+1120} $  &   $-0.18_{-0.63}^{+1.65}$      &  $<-2.18$  \\
0.53     &  0.534   & $1353_{-607}^{+1190}$  &   $-0.74_{-0.25}^{+0.35}$      &  $<-2.56$  \\
0.534    &  0.538   & $2338_{-560}^{+796}$   &   $-0.61\pm{0.17}$      &  $<-3.18$    \\
0.538    &  0.540   & $3554_{-1200}^{+2030}$ &   $-0.80\pm{0.14}$     &  $-2.51_{-2.25}^{+0.57}$   \\
0.540    &  0.542   & $2460_{-669}^{+1040}$  &   $-0.70\pm{0.17}$      &  $<-2.94$   \\
0.542    &  0.546   & $881_{-202}^{+294}$  &    $-0.36\pm{0.24}$      &  $<-3.09$     \\
0.546    &  0.548   & $418_{-112}^{+160}$  &    $-0.47_{-0.56}^{+0.93}$     &  $<-2.80$   \\
0.548    &  0.55    & $1874_{-961}^{+1493}$ &    $-0.57_{-0.28}^{+0.45}$     &  $<-2.01$     \\
0.55     &  0.554   & $6062_{-1430}^{+1790}$  &  $-0.73\pm{0.13}$     &  $<-2.31$     \\
0.554    &  0.56    & $3404_{-528}^{+674} $  &   $-0.48\pm{0.13}$      &  $<-3.12$    \\
0.56     &  0.566   & $2662_{-467}^{+583} $  &   $-0.11\pm{0.27}$      &  $<-3.57$   \\
0.566    &  0.568   & $1382_{-332}^{+700} $  &   $+0.32_{-0.62}^{+0.43}$   &  $<-2.81$    \\
0.568    &  0.570   & $1010_{-368}^{+683} $  &   $-0.66_{-0.23}^{+0.32}$     &   $<-2.54$    \\
0.570    &  0.576   & $2397_{-565}^{+736} $  &   $-0.62\pm{0.16}$     &  $<-3.10$   \\
0.576    &  0.588   & $2233_{-288}^{+335}  $ &   $-0.27\pm{0.15}$      &  $<-4.23$    \\
0.588    &  0.592   & $3485_{-484}^{+598} $  &   $0.50_{-0.37}^{+0.51}$     &  $<-3.46$   \\
0.592    &  0.596   & $3240_{-540}^{+725} $  &   $-0.19\pm{0.28}$      &  $<-3.69$   \\
0.596    &  0.600   & $1146_{-344}^{+558} $  &    $+0.06_{-0.41}^{+0.64}$     &  $<-3.17$    \\
0.600    &  0.604   & $5181_{-4210}^{+1120}$  &  $-0.91\pm{0.25}$     &  $-1.73_{-0.83}^{+0.44}$    \\
0.604    &  0.608   & $3392\pm{855} $  &   $-0.71_{-0.60}^{+1.26}$      &  $<-2.58$   \\
0.608    &  0.612   & $2606_{-1290}^{+1040}$  &  $-0.16_{-0.32}^{+0.65}$     &  $-2.66_{-1.32}^{+0.65}$ \\
0.612    &  0.616   & $1013_{-493}^{+3560} $ &   $-0.14_{-0.55}^{+0.72}$     &  $-1.71_{-0.90}^{+0.23}$\\
0.616    &  0.62    & $2101_{-927}^{+1630} $ &   $-0.72\pm{0.32}$      &  $<-2.59$\\
0.62     &  0.632   & $1887_{-942}^{+1420} $ &   $-0.63_{-0.21}^{+0.36}$      &  $<-1.77$\\
0.632    &  0.642   & $7645_{-2050}^{+3060}$  &  $-0.70\pm{0.19}$       &  $<-2.28$ \\
0.642    &  0.652   & $2443_{-1050}^{+1430}$  &   $+1.34\pm{1.20}$     &  $-2.54_{-0.84}^{+0.37}$  \\ 
0.652    &  0.658   & $4581_{-1730}^{+6740}$  &   $+0.14_{-0.34}^{+1.20}$      &  $-2.38$ \\
0.658    &  0.664   & $4629_{-1530}^{+1440}$  &   $+0.46_{-0.38}^{+1.01}$      &  $-2.60_{-0.84}^{+0.45}$ \\
0.664    &  0.67    & $6419_{-989}^{+1320} $ &   $-0.39\pm{0.16}$      &  $<-3.61$\\
0.67     &  0.678   & $3369\pm{485}$  &    $+0.05\pm{0.27}$      &  $<-3.65$\\
0.678    &  0.684   & $3814_{-889}^{+1450} $ &   $-0.44\pm{0.22}$      &  $-2.92_{-5.11}^{+0.61}$\\
0.684    &  0.692   & $3576_{-1064}^{+1418}$  &  $-0.30\pm{0.32}$      &  $-2.61_{-1.19}^{+0.54}$ \\
0.692    &  0.710   & $4445_{-1170}^{+1520}$  &  $-0.32\pm{0.27}$      &  $-2.69_{-2.96}^{+0.56}$  \\
0.710    &  0.714   & $1982_{-491}^{+593} $  &    $+0.97_{-0.76}^{+1.73}$      &  $<-2.87$\\
0.714    &  0.726   & $6646_{-1130}^{+1526}$  &  $-0.21\pm{0.25}$      &  $<-2.91$  \\
0.726    &  0.736   & $2771\pm{836} $  &   $-0.41\pm{0.20}$      &  $-2.64_{-1.51}^{+0.51}$\\
0.736    &  0.756   & $355_{-131}^{+216}$    &   $-0.37_{-0.29}^{+0.43}$      &  $-1.74\pm{0.17}$\\
0.756    &  0.850   & $159_{-53}^{+219} $    &   $-0.23_{-0.53}^{+0.64}$     &  $-1.51\pm{0.07}$\\
\hline
\end{tabular}
}
\end{center}
\end{table}

\subsubsection{$E_{\rm peak}$ distribution and evolution}
\label{Epeak_evolution_and_distribution}

\begin{figure}[h!]
\includegraphics*[width=0.5\textwidth]{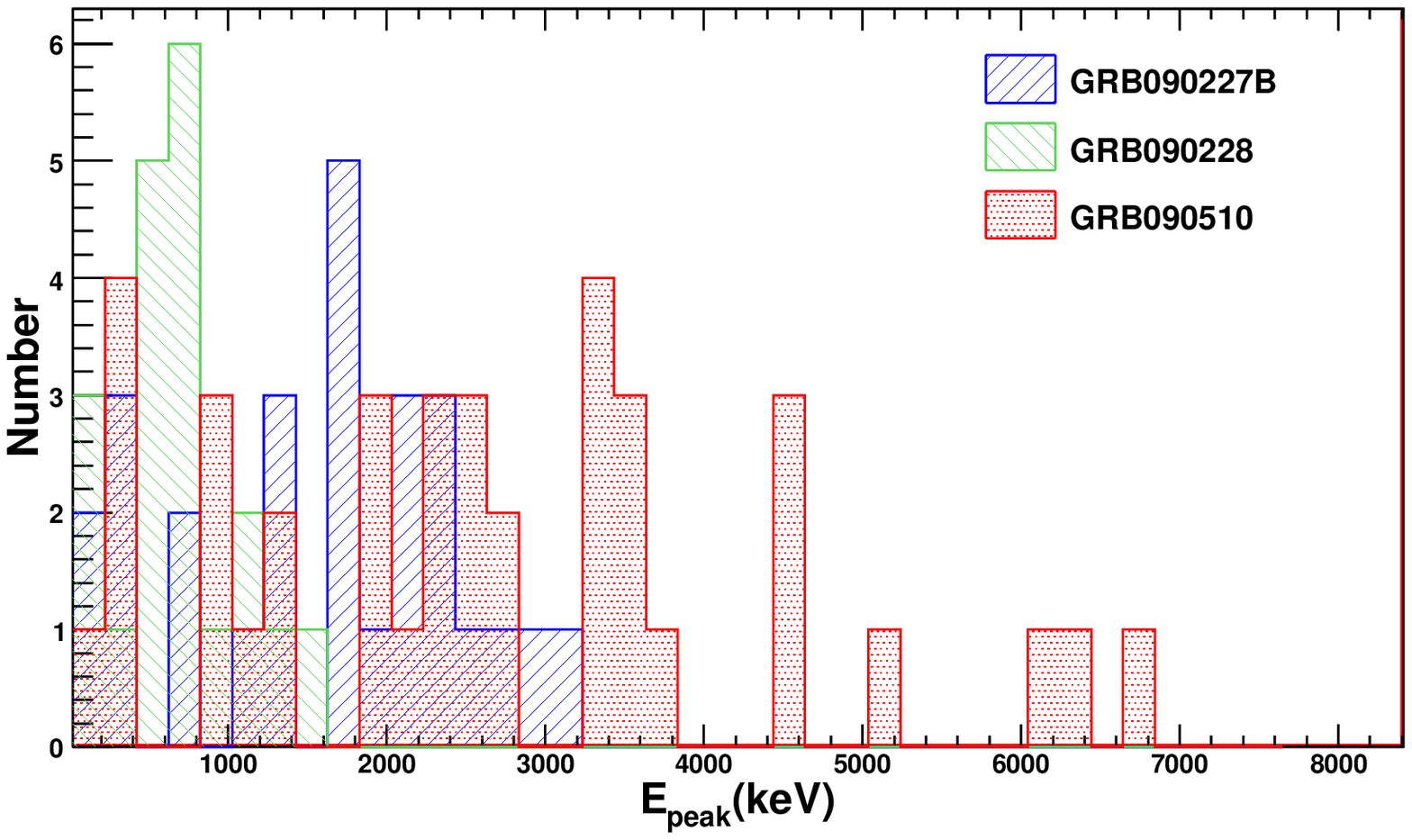}
\hspace{0.25cm}
\includegraphics*[width=0.5\textwidth]{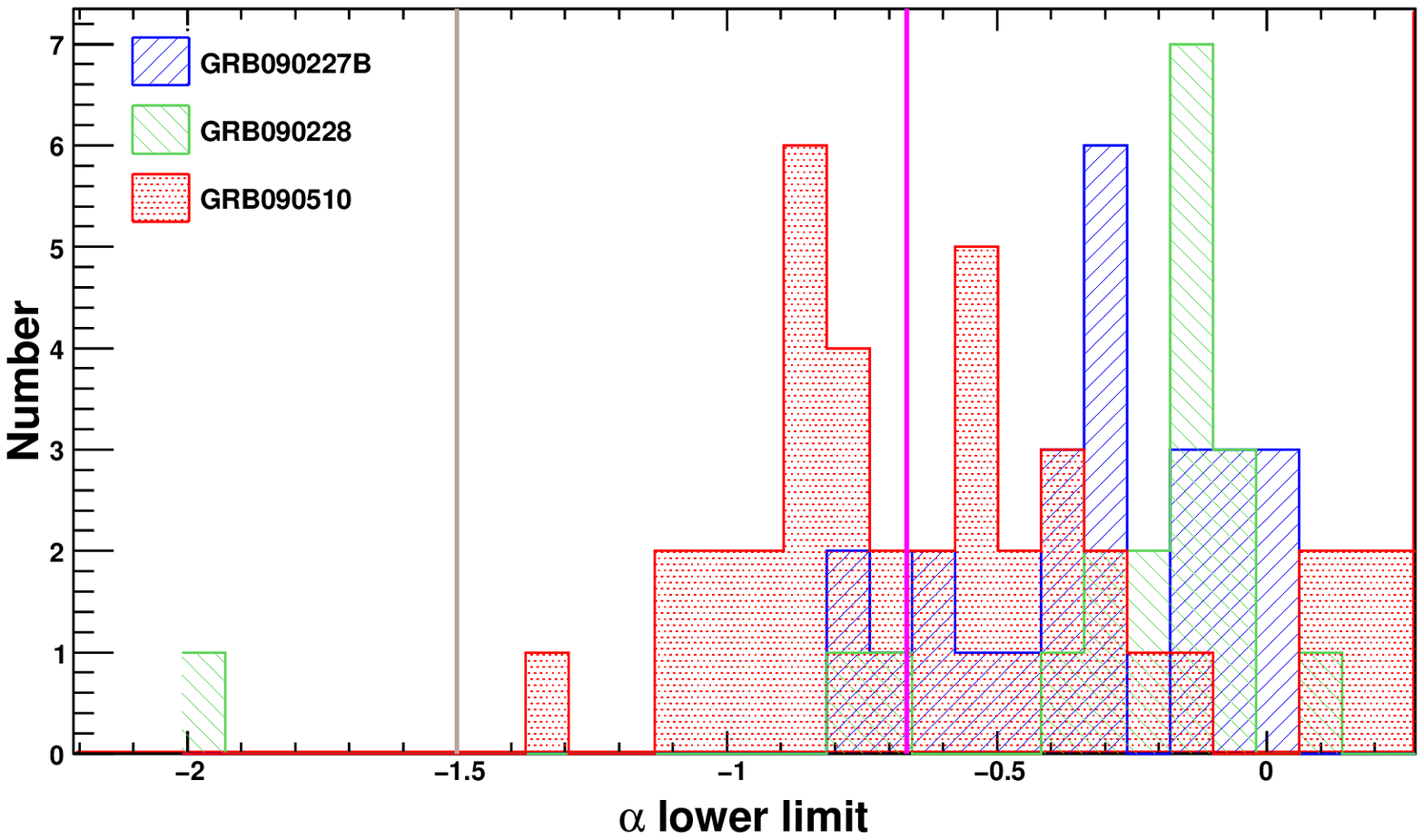}
\caption{ {\it Top:} Distribution of the peak energy, $E_{\rm peak}$, of the Band spectral function for the same 3 GRBs.
{\it Bottom:} Distribution of the 1-$\sigma$ lower limits of the low energy index $\alpha$ of the Band spectral function for GRB~$090227$B, GRB~$090228$ and GRB~$090510$. The vertical lines indicate the electron synchrotron emission limits for fast cooling (index of -3/2) and slow cooling (index of -2/3).\label{fig:alpha_Epeak_ditributions}}
\end{figure}

The distribution of the time-resolved $E_{\rm peak}$ for the three GRBs ranges over a very wide range of values (from $100\pm30$ keV to 
$3.2_{-0.7}^{+1.0}$ MeV for GRB~$090227$B, from $23\pm24$ keV to $1.5\pm0.3$ MeV for GRB~$090228$ and from $159^{+219}_{-53}$ keV to 
$7.6^{+3.0}_{-2.0}$ MeV for GRB~$090510$), indicating a strong spectral evolution in all events~(see top panel of 
figure~\ref{fig:alpha_Epeak_ditributions}). Such a strong spectral evolution is very unusual compared to what one observes in long 
bursts, where the $E_{\rm peak}$ distribution does not extend to the high energies found here~\citep{Ford:1995}.

Panels c and d of figures~\ref{fig:GRB090227B_EpeakOverLC},~\ref{fig:GRB090228_EpeakOverLC}~and~\ref{fig:GRB090510_EpeakOverLC} show the evolution of 
$E_{\rm peak}$ compared to the count-rate light-curves in two energy bands, 8 -- 200 keV and 1 -- 38 MeV, with 
the same time bins used to compute $E_{\rm peak}$. For all three GRBs, the $E_{\rm peak}$ values follow a global soft-hard-soft evolution as also seen in the high temporal resolution light-curves presented in section~\ref{section:observation}. We notice that $E_{\rm peak}$ mostly tracks the light curves, approximately correlating with the shape of the count-rate variations but not always with the intensity. We quantified the $E_{\rm peak}$ -- intensity correlation by computing the Spearman Rank Order Correlation Coeffient in the two energy ranges described above (See 
Table~\ref{table:SROCC}). We find that the correlation is most significant in the highest energy band for all three GRBs. Even with this small sample, the independent correlation measurements increase the robustness of our results.

\begin{table}[h!]
\caption{\label{table:SROCC} Spearman Rank Order Correlation Coefficients ($\rho$) and associated probability computed between the raw counts light curves in the 8--200 keV and 1--38 MeV energy bands, and the $E_{\rm peak}$ values plotted at the bottom of figures~\ref{fig:GRB090227B_EpeakOverLC}~to~\ref{fig:GRB090510_EpeakOverLC}.}
\hspace{-0.5cm}
{\tiny
\begin{tabular}{|l|c|c|c|c|c|}
\hline
\multicolumn{2}{|c|}{Energy bands :} & \multicolumn{2}{|c|}{8 -- 200 keV} & \multicolumn{2}{|c|}{1 -- 38 MeV} \\
\hline
\multicolumn{1}{|c|}{Name } & \multicolumn{1}{|c|}{Number of data points} & \multicolumn{1}{|c|}{$\rho$} & \multicolumn{1}{|c|}{probability} & \multicolumn{1}{|c|}{$\rho$} & \multicolumn{1}{|c|}{probability} \\
\hline
GRB~$090227$B & 27 &  0.354 & 0.076 & 0.452 & 0.021 \\
GRB~$090228$  & 20 &  0.498 & 0.030 & 0.666 & 0.002 \\
GRB~$090510$  & 39 & -0.014 & 0.935 & 0.371 & 0.022 \\
\hline
\end{tabular}
}
\end{table}

\begin{figure}
\includegraphics*[width=0.5\textwidth]{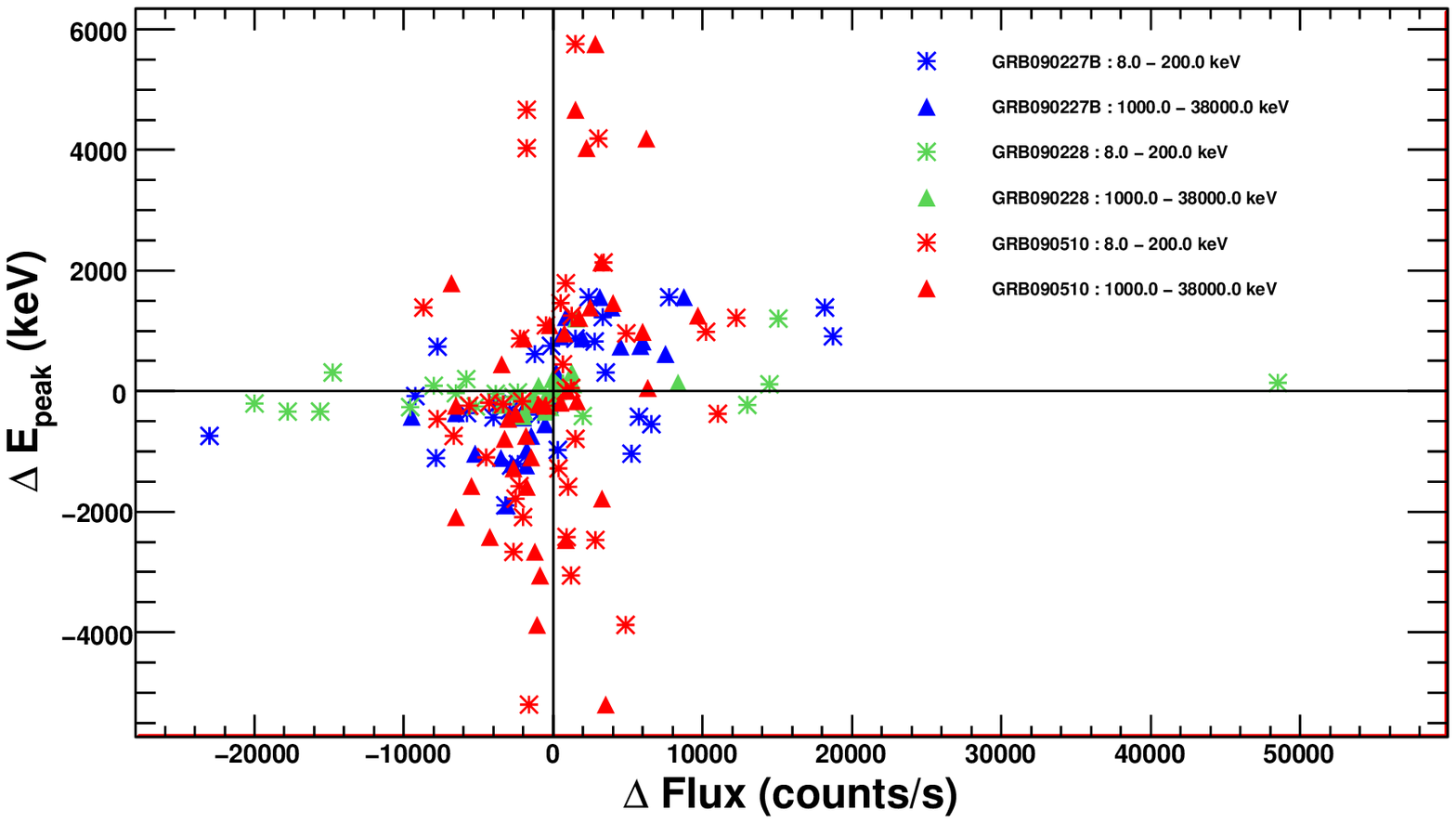}
\includegraphics*[width=0.5\textwidth]{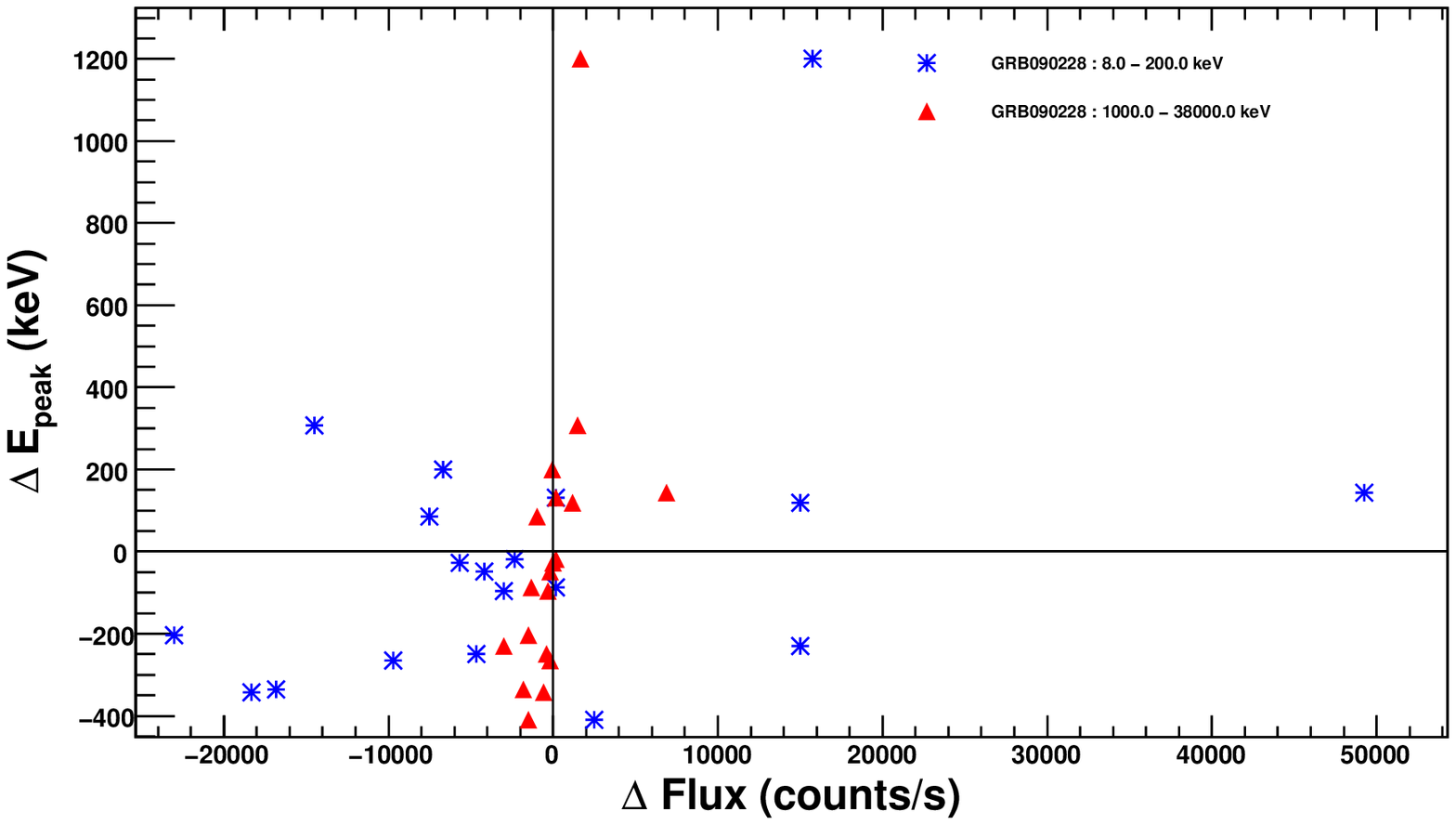}
\caption{{\it Top:} Variation of the peak energy $E_{\rm peak}$ versus the count rate variation for two energy bands 8-200 keV (stars) and 1-38 MeV (triangles) for GRB~$090227$B, GRB~$090228$ and GRB~$090510$ (red, blue, and green, respectively). {\it Bottom:} zoom in on GRB~$090228$ only (8-200 keV: blue stars; 1-38 MeV: red triangles). \label{fig:DeltaEpeakVSDeltaFlux_ditribution}}
\end{figure}

The $E_{\rm peak}$-intensity correlation is a well known characteristic of long GRBs~\citep{Ford:1995,Golenetskii:1983}, where frequently the 
peak in $E_{\rm peak}$ leads the peak in intensity~\citep{Ford:1995}. This pattern seems weak or absent in our short events, possibly because the low statistics in such short time intervals make it necessary to combine too many bins to see this effect.
To study the simultaneous changes in $E_{\rm peak}$ values and intensities of these events we computed and displayed the derivatives of their values (we cannot use cross-correlation techniques as we are limited by the small number of data points). Figure~\ref{fig:DeltaEpeakVSDeltaFlux_ditribution} shows a scatter plot of these derivatives for all three bursts. Points in the +/+ and -/- quadrants indicate a correlated increase or decrease of the two data sets, respectively. A positive correlation is evident, especially in the 1--38 MeV light curves, indicating that increasing (decreasing) flux correlates with increasing (decreasing) $E_{\rm peak}$. 


\subsubsection{Low and high-energy power-law indices}

The bottom panel of Figure~\ref{fig:alpha_Epeak_ditributions} shows a histogram of the $1\sigma$ lower-limits of the low-energy index 
$\alpha$. The $\alpha$ values are exceptionally hard, increasing the discrepancy with the predictions of synchrotron emission even if they don't rule out this emission mechanism as the main process to explain the prompt emission, as we will discuss 
in section~\ref{section:interpretation}. The time-resolved values of $\alpha$ are even harder than the time-integrated 
ones (also described in section~\ref{section:standard_parameters}) and most of the values exceed $-\frac{2}{3}$, the electron slow-cooling limit.

This very hard distribution for both $\alpha$ and $E_{\rm peak}$ confirms the results presented in~\citet{Paciesas:2003}, which describes 
the short GRBs as harder with both higher $E_{\rm peak}$ values and harder $\alpha$, and does not support the explanation offered by~\citet{Ghirlanda:2004,Ghirlanda:2009b}.

It is also remarkable that the time-resolved $\beta$ values (Tables~\ref{table:time_resolved_spectroscopy_GRB090227B} to~\ref{table:time_resolved_spectroscopy_GRB090510}) are mostly very soft according to their 1$\sigma$ upper limits: from $<-4.30$ to $<-1.74$ for GRB~$090227$B, from $<-3.97$ to $<-2.11$ for GRB~$090228$ and 
from $<-4.23$ to $<-1.29$ for GRB~$090510$. The variations of this index between the hardest well-constrained values and the softest 1$\sigma$ upper 
limits also indicate strong evolution: from $<-4.30$ to $-1.96^{+0.21}_{+0.31}$ for GRB~$090227$B, from $<-3.97$ 
to $-2.39^{+0.33}_{-0.60}$ for GRB~$090228$ and from $<-4.23$ to $-1.73^{+0.44}_{-0.83}$ for GRB~$090510$.

\section{Interpretation}
\label{section:interpretation}


The prompt emission from short and long GRBs shows remarkably similar properties. The main difference appears in the hardness-duration diagram : short GRBs are harder \citep{Kouveliotou:1993}. It is therefore usually believed that short and long gamma-ray bursts are produced by different progenitors leading to the same succession of events: formation of a compact central engine and relativistic ejection in the form of collimated jets \citep[e.g.][for a review]{Nakar:2007}. In this picture, the prompt emission is produced by the same mechanism for all GRBs and the observed differences between short and long bursts have to be explained by different initial conditions (energetics and lifetime of the central engine, source environment, etc.). Due to the short variability timescale observed in GRBs, the prompt emission has very likely an internal origin \citep{Sari:1997}, which means that it is not related to the deceleration of the jet by the ambient medium. This scenario favors a greater similarity in the prompt properties of short and long GRBs than for their afterglows. There are then three possible energy reservoirs from which the prompt radiated energy can be extracted, in a region located above the photosphere of the relativistic outflow and below its deceleration radius: (i) the internal (thermal) energy, that can be extracted at the photosphere \citep[e.g.][for recent discussions of this mechanism]{Peer:2008,Beloborodov:2009} ; (ii) the kinetic energy, that can be extracted via (internal) shock waves \citep{Rees:1994}; (iii) the magnetic energy -- if the outflow is strongly magnetized -- which can be extracted via magnetic reconnection \citep[e.g.][]{Spruit:2001,Lyutikov:2003}. In the latter case, the physics of the magnetic reconnection in relativistic outflows is still far from being understood and it is difficult to predict precisely the spectrum of the expected radiation. For this reason, we will not discuss it further. In addition to the mode of energy extraction, one also needs to understand the distribution of the emitting particles, which constrains acceleration mechanisms, and the nature of the dominant radiative processes.\\

\subsection{Why are short GRBs harder?} 

The results presented here offer new clues to identify which mechanism is at work during the prompt emission of short GRBs.
In the three bright short GRBs studied here, the broad spectral range of the GBM instrument on board {\it FGST} allows us 
to demonstrate clearly that the origin of their hardness is mainly due to very high peak energies, well above the usual range observed in long GRBs. This is naturally expected in internal shocks \citep{Daigne:1998}.
Precisely, the observed isotropic peak energy, $E_\mathrm{p,obs}^\mathrm{IS}$, associated with a collision scales as \citep{Barraud:2005}
\begin{equation}
\label{equation1}
E_\mathrm{p,obs}^\mathrm{IS} \propto \frac{\epsilon_\mathrm{B}^{1/2}\epsilon_\mathrm{e}^2}{\zeta^2}\phi\left(\kappa\right) \frac{\dot{E}^{1/2}}{\bar{\Gamma}^2\tau }\ ,
\end{equation} 
where $\dot{E}$ and $\bar{\Gamma}$ are the isotropic equivalent kinetic energy flux and mean Lorentz factor in the outflow and $\tau$ is the variability timescale during the relativistic ejection. The contrast $\kappa$ is defined as the ratio between the maximum and the minimum Lorentz factor in the outflow and the function $\phi(\kappa)$ is steadily increasing \citep[see][]{Barraud:2005}. The microphysical parameters $\epsilon_\mathrm{e}$ and $\epsilon_\mathrm{B}$ are the fraction of the energy dissipated in internal shocks that are injected respectively in relativistic electrons and in the magnetic field, and $\zeta$ is the fraction of electrons that are accelerated. The isotropic equivalent gamma-ray luminosities in the small sample of short bursts that have a measured redshift are similar as for long GRBs (see \citet{Nakar:2007} and Fig.~3(b) in \citet{Zhang:2009}). Therefore, one can assume similar kinetic energy fluxes $\dot{E}$ in short and long GRBs. On the other hand,
we show here that the lightcurves of the three short GRBs imply that the variability timescales
are contracted, compared to long GRBs, during the relativistic ejection, which is most probably caused by the differences in the central engine. 
If all other properties are fixed (distribution of the Lorentz factor, microphysical parameters), this leads to higher peak energies as $E_\mathrm{p,obs}^\mathrm{IS}\propto \tau^{-1}$ and gives a natural explanation for the observed hardness-duration relation in GRBs. This relation should only be a general tendency, since variations in the Lorentz factor ($\bar{\Gamma}$, $\kappa$) or the microphysical parameters ($\zeta$, $\epsilon_\mathrm{e}$, $\epsilon_\mathrm{B}$) from one burst to another will lead to some dispersion according to the equation above.

More generally, for many possible radiative processes, the peak energy in the comoving frame should scale as $E_\mathrm{p,com}^\mathrm{IS} \propto \rho_*^x \epsilon_*^y$ where $\rho_*$ and $\epsilon_*$ are the comoving density and specific energy density in the emitting shocked region. This leads to the following general expression of the observed peak energy (Barraud et al. 2005) : 
\begin{equation}
\label{equationx}
E_\mathrm{p,obs}^\mathrm{IS} \propto \Phi_{xy}(\kappa) \frac{\dot{E}^{x}}{\bar{\Gamma}^{6x-1}\tau^{2x}}\, , 
\end{equation}
where $\Phi_{xy}$ is now a function depending on the two exponents $x$ and $y$. Synchrotron radiation with constant microphysics parameters corresponds to $x=1/2$ and $y=5/2$, which leads to equation~\ref{equation1} above. Inverse Compton scattering in Thomson regime corresponds to $x=1/2$ and $y=9/2$, which leads to the same dependance of the peak energy on the duration. The fact that the observed peak energy increases in internal shocks when the timescale decreases remains valid as long as $x>0$, e.g. as long as  the peak energy in the comoving frame decreases with the density, which seems a reasonable assumption. This is for instance the case for the jitter radiation \citep{Medvedev:2009} if the correlation scale of the magnetic field increases when the density decreases.
While the hardness-duration relation seems a robust feature of the internal shock model, the slope of the predicted relation depends, however, on the details of the dominant radiative process and/or the acceleration mechanism in mildly relative internal shocks, as well as the dispersion due to the distribution of the other parameters (Lorentz factor, kinetic energy flux, etc).

In the internal shock model, it is also expected that the peak energy should track the lightcurve. If a given pulse is associated with the propagation of a shock wave within the outflow, the magnetic field decreases when the radius increases, resulting in a decay of the peak energy in the decay phase of the pulse  \citep{Daigne:1998,Daigne:2003}. \\


In photospheric models, the observed peak energy is related to the temperature of the fireball at the photospheric radius and should therefore scale as 
\begin{equation}
\label{equationFin}
E_\mathrm{p,obs}^{TH}\propto \frac{\bar{\Gamma}^{8/3} r_\mathrm{0}^{1/6}}{\dot{E}^{5/12}},
\end{equation}
where $r_0$ is the radius where the initial relativistic ejection takes place. It appears then  less natural to expect short GRBs to have higher peak energies in this scenario, as the shorter timescales in short GRBs would imply smaller radii $r_0$. Higher peak energies would be obtained if -- due to differences between the  central engines -- outflows in short GRBs have systematically higher Lorentz factors. As only lower limits on $\bar{\Gamma}$ have been obtained so far for {\it FGST} GRBs~\citep{Abdo:2009:GRB090902B,Abdo:2009,Abdo:2009:GRB080916C}, this is still a possibility. However, as these lower limits are already quite large, this could lead to a severe constraint for models of jet acceleration in central engines of short GRBs.\\

\subsection{On the low energy power law index, $\alpha$}

The simple scenario where the prompt emission in (short and long) GRBs is due to emission of relativistic electrons 
accelerated in internal shocks within a relativistic outflow, offers a consistent framework to explain many observed features: high variability in lightcurves, spectral evolution, time lags, etc. In addition, compared to the photospheric model, it seems to give a natural explanation for the observed hardness-duration relation, which is a fundamental observation in the comparison between the two classes of bursts. In the internal shock model, short GRBs have higher peak energies because of shorter variability timescales, which seems consistent with the GBM observations discussed in this paper. Most of these properties are due to the dynamics of internal shocks and are therefore robust predictions, found for different assumptions regarding the dominant radiative process and the microphysics in mildly relativistic shocks. The two most discussed possibilities are synchrotron radiation or synchrotron self-Compton (SSC). The fact that strong components in the GeV range, dominating the total energy release, are not a common feature in FGST bursts favors the synchrotron scenario for long bursts \citep{Bosnjak:2009,Piran:2009}. There is, however, a remaining worrying problem that appears once again in the current analysis: the observed low-energy photon index, $\alpha$, is too steep to be easily explained by synchrotron emission. The expected slope in the fast-cooling regime is $\alpha=-3/2$ \citep{Ghisellini:2000}. Inverse Compton scattering in the Klein-Nishina regime can affect the electron cooling and steepen the synchrotron spectrum up to $\alpha=-1$ \citep{Derishev:2001,Bosnjak:2009,Nakar:2009}. Synchrotron radiation in the slow-cooling regime allows to reach $\alpha=-2/3$ but leads to a problem with the radiative efficiency. In the present sample of three bright short bursts, the measured values of $\alpha$ are almost always above this limit $\alpha=-2/3$, which is another difference with long GRBs showing a mean value close to $\alpha=-1$ \citep{Preece:2000}. These observed values of $\alpha$ are really challenging for the synchrotron process. Jitter radiation can produce such flat slopes \citep{Medvedev:2009}. It remains to be confirmed, however, that it is a viable process for physical conditions expected in internal shocks \citep{Medvedev:2009}.
The SSC mechanism can also lead to steeper values of $\alpha$ but it should then be understood why the additional component in the LAT range are not brighter. Photospheric models can also show very steep low-energy spectral slopes but on the other hand, they do not propose natural explanations for the hardness-duration relation and more generally the observed spectral evolution.\\

\subsection{On the high energy power law index, $\beta$}

The broad spectral range of GBM in principle allows investigation of the values of the high-energy slope $\beta$ but
the observed high values of the peak energies makes this measurement difficult in the three short GRBs presented here. In many cases, a power-law at high energies is not statistically preferred to an exponential cutoff. The time-resolved spectral analysis presented here allows us, however, to constrain the value of $\beta$ in most time bins.  The measured values of $\beta$ are almost always compatible with $\beta < -2$ which shows that the main component is recovered, without any need for an additional break at higher energies. If the dominant process is synchrotron radiation from shock-accelerated electrons, such measurements can be translated into constraints on acceleration theory in mildly\footnote{{\footnotesize Unlike the external shock, which is a relativistic collision between the relativistic outflow and the resting ambient medium, internal shocks occur between density fronts with mildly-relativistic relative-velocities within the outflow.}} relativistic shocks.  This regime is not well covered by present simulations which usually focus on the ultra-relativistic regimes expected in afterglows~\citep{Spitkovsky:2008a}.
The fact that $\beta$ is variable within a GRB would be an indication that there is no universal slope for the distribution of shock-accelerated electrons in the mildly relativistic regime. In addition, the observation that $\beta$ is very steep would also indicate that the distribution of shock-accelerated electrons is very peaked at low Lorentz factors, as the predicted slope $-p=2\left(\beta+1\right)$ lies in the typical range~$-6$~to~$-2$ for $\beta=-4$~to~$-2$. This could also simply indicate that the distribution of electrons is more complex than a single power-law~\citep{Spitkovsky:2008b,Martins:2009}.  The true distribution of electrons could however be masked by other effects that could steepen $\beta$, especially if the real spectral shape at high energies for the main component is a power-law of slope $\beta$ above $E_\mathrm{peak}$ followed by a cutoff above $E_\mathrm{cut} > E_\mathrm{peak}$. The measured slope could appear as steeper than $\beta$ if $E_\mathrm{cut}$ is not at too high energy. Several physical processes could be responsible for such a cutoff: for instance the existence of a maximum Lorentz factor for shock accelerated electrons or $\gamma\gamma$ annihilation above the threshold for pair production. The resulting spectral shape could be rather complex. It is beyond the scope of this paper to model such processes in detail but the broad spectral range of GBM, especially when combined with the LAT, clearly offers a new opportunity to make progress in this field in the future.\\

\subsection{On the additional spectral component}

In addition to the main component which is well fitted by the Band GRB function, there is evidence of an additional power-law component, dominant at low energy, and -- at least in the case of GRB~$090510$ and GRB~$090227$B -- an additional component at high energy which could be related \citep{Abdo:2009,Ackermann:2010:GRB090510}.  This suggests that the radiative processes at work are more complicated than was predicted in the simplest model, possibly mixing several distinct components with varying relative weights. 

The difficulty in identifying the mechanism responsible for the main spectral component that peaks in the GBM energy range makes the physical modeling of this additional component even more challenging.
In scenarios where the soft gamma-rays are produced at the photosphere, the additional component could be the signature of synchrotron radiation from fast-cooling electrons accelerated in internal shocks 
\citep[e.g.][]{Ryde:2010}. It implies, however, that the peak $h\nu_\mathrm{m}$ of the synchrotron radiation is at  high energy (LAT range), which leads to extreme constraints on the magnetic field $B$ and the electron Lorentz factor $\Gamma_\mathrm{m}$ :
\begin{equation}
B \Gamma_\mathrm{m}^2 \simeq 6\times 10^{13}\, \left(\frac{\bar{\Gamma}}{1000}\right)^{-1} \left(\frac{h \nu_\mathrm{m,obs}}{1\, \mathrm{GeV}}\right)\, G
\end{equation}
On the other hand, if internal shocks are responsible for the soft gamma-ray emission, several mechanisms can produce additional components at low and/or at high energy.

In addition to the leptonic synchrotron component, a weak inverse Compton component can be present at high energy \citep{Bosnjak:2009}, as well as a hadronic component \citep{Asano:2009}. The lack of identification of a clear cutoff in the high-energy spectrum of GRBs detected by the LAT makes it difficult to evaluate if a strong $\gamma\gamma$ annihilation occurs at high energies even if recent observations indicate such a cutoff \citep{Abdo:2009:GRB090902B, Ackermann:2010:GRB090926}. If this is the case, the new component at low energies could be associated with the radiation of the produced electron-positron pairs and would then be related to the high-energy component. Alternatively, even in the internal shock model,
 photospheric emission is usually also expected, and could be superimposed on the synchrotron radiation~\citep{Daigne:2002}, which can again lead to an additional component in the spectrum, preferentially in the GBM range or below.  Note that in the latter case, different central engines and acceleration mechanisms for the outflow could lead to different intensities of the photospheric component in short and hard GRBs. We cannot assess here the viability of these models, but our analysis does illustrate that introducing more realistic and complex physics is probably now necessary to model the spectral shape observed in GRBs. Finally, one should keep in mind that our study is based on three short bursts only, selected by their high brightness and that they are not necessarily fully representative of the short GRB population.

\section{Conclusions}

With GBM, it has been possible for the first time to perform time resolved spectroscopy of short GRBs with a resolution down to 2 ms. The time-integrated spectra of the three GRBs analysed here are best fit with the sum of a power law with an exponential cutoff and an additional power law with indices similar for the three GRBs, clustering around $-$1.5. This additional power law indicates a deviation from the standard Band function usually used to fit GRB prompt emission spectra in the keV-MeV energy range and can be interpreted as an additional component. We find that this extra component overpowers the Band function at low energies (below a few tens of keV), and at high energies (above a few MeV), and  could be a combination of various mechanisms as explained in Section~ \ref{section:interpretation}. How the low and high energy excesses are related together is still not clear yet.


The spectral parameters obtained from the fits of the integrated spectra are much harder than usually observed in long GRBs for both the low energy index $\alpha$ ($>-0.43$) and the $E_{\rm peak}$ (up to about 4 MeV) values. When a Band+PL model is used, the high-energy spectral index below $-$2.90 indicates a soft spectrum above the break energy. The time-resolved and time-integrated low energy indices nearly always violate both the electron slow and fast cooling predicted by the synchrotron models and additional emission processes may be required to explain the data.

Based in the observed pulse structures, we posit that the light curves of short GRBs are similar to the ones of long GRBs but contracted in time. Future observations that probe even shorter timescales in both types of GRBs will test this hypothesis. The fine-time resolved spectroscopy shows that the short GRB $E_{\rm peak}$ values are stretched up to higher energies compared to long burst ones. The time-resolved $E_{\rm peak}$ values can vary from few tens of keV up to several MeV in a time scale as short as a few hundreds of ms. $E_{\rm peak}$ mostly tracks the light curve evolution in a way that is similar to long bursts. A convincing hardness-intensity correlation has been measured in these three GRBs between $E_{\rm peak}$ and the light curves above 1 MeV. A more general correlation between the signs of the derivatives of the $E_{\rm peak}$ curves and the light curves seems to be present for these three GRBs. These results favor the scenario where the same physical mechanism is at work in the prompt phase of short and long GRBs. The fact that short GRBs are harder would then be mainly due to their shorter timescales, probably a signature of different central engines. The internal shock model naturally leads to the expected hardness-duration relation for a broad range of radiative processes whereas such a relation appears less natural for other energy extraction mechanisms.


\section{Acknowledgments}

We thank the referee and the editor for their useful comments, which increased the quality of the paper.

The GBM project is supported by the German Bundesministerium f\"ur Wirtschaft und Technologie (BMWi) via the Deutsches Zentrum f\"ur Luft- und Raumfahrt (DLR) under the contract numbers 50 QV 0301 and 50 OG 0502.

A.J.v.d.H. was supported by an appointment to the NASA Postdoctoral Program at the MSFC, administered by Oak Ridge Associated Universities through a contract with NASA.

S.F. acknowledges the support of the Irish Research Council for Science, Engineering and Technology, cofunded by Marie Curie Actions under FP7.


\FloatBarrier
\bibliographystyle{bibstyles/astron}
\bibliography{guiriec}

\end{document}